\documentclass[12pt, draftclsnofoot, onecolumn]{IEEEtran}
\usepackage{dblfloatfix}  
\usepackage{amsfonts}
\usepackage[T1]{fontenc}
\usepackage{setspace}
\usepackage{cite}
\usepackage[latin1]{inputenc}
\usepackage{epsf}\epsfverbosetrue
\usepackage{graphics,epsfig,color}
\usepackage{graphicx,epsfig}
\usepackage{epsfig}
\usepackage{multirow}
\usepackage{slashbox}
\usepackage{alltt}
\usepackage{subfigure}
\usepackage{color}
 \usepackage{float} 
\usepackage{url}
\usepackage{graphicx}
\usepackage{verbatim} 
\usepackage{latexsym} 
\usepackage{amsmath}
\usepackage{color}
\usepackage{wrapfig}
\usepackage{algorithm}
\usepackage{eso-pic}
\usepackage{fix-cm}
\usepackage{color}
\usepackage{wrapfig}
\usepackage{algorithm}
\usepackage{layout}
\usepackage{amsmath}
\usepackage{textcomp}
\usepackage{multirow}
\usepackage{soul,xcolor}
\usepackage{amsmath}
\usepackage{verbatim}
\usepackage{tikz}
\begin{document}
\setstcolor{red}
\title{Effective Capacity  of NOMA with Finite Blocklength for  Low-Latency Communications}
\author{Muhammad Amjad, Leila Musavian, and Sonia A\"{i}ssa
\thanks{M. Amjad and L. Musavian are with the School of Computer Science and Electronic Engineering, University of Essex, CO4 3SQ, UK (Emails:\{m.amjad, leila.musavian\}@essex.ac.uk). S. A\"{i}ssa is with the Institut National de la Recherche Scientifique (INRS-EMT), University of Quebec, Montreal, QC, H5A 1K6, Canada (Email: aissa@emt.inrs.ca).}
\thanks{Part of this work has been submitted to IEEE International Symposium on Personal, Indoor and Mobile Radio Communications (PIMRC), 2020.}
}
\maketitle
\begin{abstract}
\noindent In this paper, we investigate the link-layer rate of a two-users non-orthogonal multiple access (NOMA) network in finite blocklength (short packet communications) regime, where the two users are paired from a set of $V$ users. The overall reliability consists of the transmission error probability and the queuing delay violation probability. The performance of the two-users NOMA network in finite blocklength is verified for achieving latency and reliability, using the effective capacity (EC) framework. Specifically, we derive closed-form expressions for the EC of the two-users NOMA network in finite blocklength regime, considering transmissions over Rayleigh fading channels. We also study a multiuser NOMA network and derive the total EC of two-users NOMA subsets, and show that the NOMA set with users having distinct channel conditions achieve maximum total EC. Focusing on a two-users   NOMA network, we study the impact of the transmit signal-to-noise ratio, delay exponent, and transmission error probability, on the achievable EC of each user. The analysis shows that when the delay exponent is high, the delay violation probability does not improve below a certain value due to the dominant factor of the transmission error probability. The accuracy of the proposed closed-form expressions for the individual EC of a two-users NOMA is verified using the Monte-Carlo simulations.
\end{abstract}
\newpage
\section{Introduction}
\label{sec:in}
\noindent Major use cases of the 5th generation (5G) and beyond 5G (B5G) cellular communications require the provision of ultra reliable and low-latency communications (URLLC).  Non-orthogonal multiple access (NOMA) in conjunction with finite blocklength (short packet communications)\footnote{The terms finite blocklength and short packet communications will be used alternatively throughout the paper.} is considered as key enabler for URLLC \cite{R178,R136}. In fact, NOMA has gained much attention in academia and industry due to its potential to achieve higher throughput, massive connectivity, low latency, and higher reliability in favorable circumstances, as compared to its orthogonal multiple access (OMA) counterpart \cite{R180,R181}. NOMA with finite blocklength follows the basic operation of traditional NOMA, with superposition coding (SC) at the transmitter and successive interference cancellation (SIC) at the receiver. However, the conventional Shannon formula to approximate the attainable rate (with almost no errors) is not applicable when considering short packets in the communications \cite{R153}.

To achieve latency as low as $1$ms, and reliability as high as $99.999\%$, communications in finite blocklength regime is very promising \cite{R016,R153,R086}. In the leading work on finite blocklength \cite{R139}, the achievable rate of finite blocklength communication link constrained by a given error probability was investigated in additive white Gaussian noise (AWGN) channels. In \cite{R139}, the blocklength was taken as small as 100 bits, and it was shown that the maximal achievable rate with finite blocklength could not be approximated with the Shannon formula. In the same work, a penalty factor which is the function of the channel dispersion and error probability was introduced to obtain the achievable rate in finite blocklength regime. The study work was further extended for the case of Rayleigh block fading channels in \cite{R047}, wherein a trade-off between reliability, latency, and throughput in finite blocklength regime was investigated to establish the importance of short packet communications for low latency. Furthermore, upper and lower bounds on the received  signal-to-noise ratio (SNR) while considering the finite blocklength for a given error probability were obtained in \cite{R047}.

To analyze the suitability of short packet communications for low latency, the effective capacity (EC) framework was used in \cite{R136,R157}. EC is the dual concept of effective bandwidth, and is used to find the maximum arrival rate for a given service rate while satisfying a certain delay constraint \cite{R147,R158}. The performance of short packet communications to achieve URLLC was also investigated in \cite{R154,R156} using the EC concept. For example, in \cite{R154}, the performance of a point-to-point network under latency constraint while considering finite blocklength transmission  was investigated. In that work, three different transmission strategies (fixed-rate, variable-rate, and variable-power) were studied with focus on short packet communications. Later, the closed-form expression for the achievable EC with short packet communications Rayleigh fading channels for machine-type communications was found in \cite{R156}. The latter work solely considered the ultra-reliable communications (URC) use case, but did not consider URLLC. The performance of short packet communications for achieving low latency were investigated with the EC concept in \cite{R157}.

On the other hand, NOMA combined with finite blocklength is considered as an enabling technology for low latency communications. In this regard, performance of NOMA with finite blocklength was investigated in \cite{R140}, which showed the amount of physical-layer transmission latency that NOMA with finite blocklength can reduce under reliability constraint as compared to OMA. In \cite{R140}, a closed-form expression for the block error rate of a two-user NOMA was derived and validated with simulations. Authors in \cite{R166} considered  NOMA with short packet communications, and focused on the trade-off between decoding error probability, transmission rate, and blocklength. More specifically, in \cite{R166}, the challenges associated with the SIC and transmission rate while using finite blocklength were highlighted. The latency performance of NOMA in finite blocklength regime as compared to its OMA counterpart was investigated in \cite{R176}. This work showed the improved performance of NOMA in terms of throughput and reducing latency as compared to OMA with finite blocklength. Another work on the comparative analysis of NOMA and OMA in short blocklength regime under reliability and latency constraints was done in \cite{R177}. The latter work was focused on energy efficient transmission with NOMA, and showed improved performance as compared to OMA. On the other hand, a detailed statistical delay analysis of NOMA using EC was conducted in \cite{R159}, including closed-form expressions for the achievable EC of a two-users NOMA when the users are chosen from a set of $V$ users. The work in \cite{R159}, however, did not consider the latency performance of NOMA with a short packet communications. NOMA with short packet communications was investigated with the concept of effective bandwidth in \cite{R160}, where the required SNR for a given delay exponent and transmission error probability was obtained. Link-layer rate of two-users NOMA and OMA in finite blocklength was investigated in \cite{R183}. This work showed the improved performance of NOMA as compared to OMA when the delay exponent is loose. However, detailed delay analysis of NOMA in finite blocklength, i.e., multiple NOMA users and the impact of delay exponent on queueing delay violation probability, is yet to be investigated.

In this paper, we derive the achievable EC of a two-users NOMA network, when the users are chosen from a set of $V$ users in the cell, with finite blocklength regime to investigate the low-latency communications. The impact of a given transmission error probability, delay exponent, and transmit SNR, on the achievable EC with finite blocklength is investigated in detail. The major contributions of this paper can be summarized as follows:
\begin{itemize}
 \item The achievable EC of downlink  two-users (out of $V$ users) NOMA, and also a multiuser NOMA network, with finite blocklength, are derived.
 \item Closed-form expressions for the achievable total EC of two-users NOMA subset, as well as the achievable individual EC of each user, in finite blocklength regime, are derived in Section \ref{sec:ec-noma-short}. These expressions are validated using using Monte-Carlo simulations in Section \ref{sec:numerical-results}.
 \item Realizing the complexity of the proposed closed-form expressions for the two-users NOMA, we also derive simplified closed-form expressions to approximate the EC of the two-users NOMA network at high transmit SNRs.
 \item Total EC of multiple NOMA pairs in finite blocklength regime is also investigated by taking into consideration the different pairing sets of multiple users. These findings show that a NOMA set with users having more distinct channel conditions achieve a higher total EC as compared to one where users have less distinct channel conditions.
 \item Focusing on the  two-users NOMA network in finite blocklength, the impact of delay exponent, queuing delay violation probability, transmission error probability, and transmit SNR, on the achievable EC of the strong and weak users is analyzed through simulations. In particular, it is shown that when the delay exponent becomes stringent, the  queueing delay violation probability cannot be reduced below a certain value due to the dominance effect of transmission error probability.
\end{itemize}
The remainder of this paper is organized as follows. First, the system model is discussed in Section \ref{sec:system-model}. Concepts related to the theory of EC are presented in Section \ref{sec:theory-ec}. Then Section \ref{sec:ec-noma-short} provides the achievable EC of two-users NOMA with finite blocklength. Numerical results with their insights are investigated in Section \ref{sec:numerical-results}, and the paper is concluded in Section \ref{sec:conclusion}.
\section{System Model}
\label{sec:system-model}
\begin{figure*}
\centering
  \includegraphics[width=\linewidth]{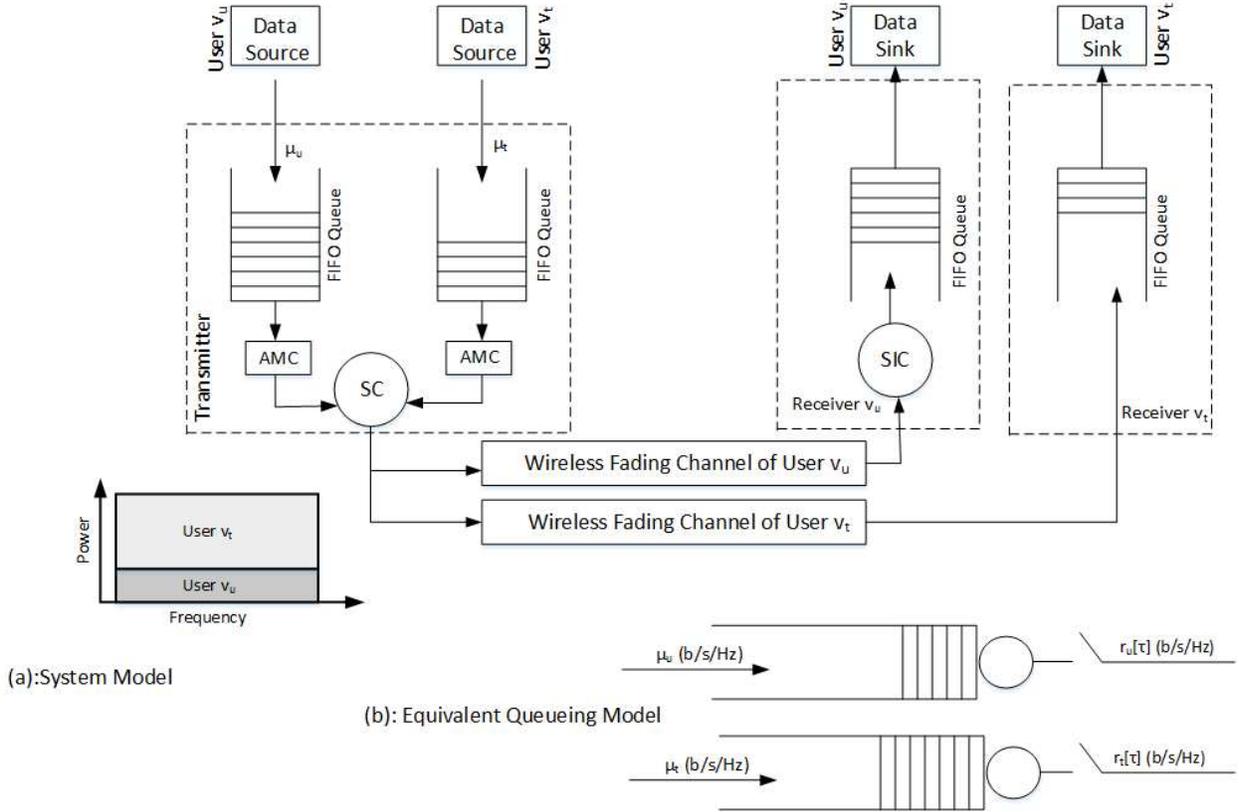}
\caption{Two-user NOMA operation with finite blocklength with their respective queues: (a) describes the system model with two queues at the BS with their respective receivers, and (b) depicts the equivalent queueing model with the arrival rate and service rate.}
\label{fig:sys2}
\end{figure*}

In this work, we consider power-domain downlink NOMA network with short packet communications. The network consists of one base station (BS) and $V$ single-antenna users. The upper-layer packets of each user are assembled into frames, then stored at the transmit buffer at the BS, and later transmitted over the wireless channel as bit streams. We assume that each user is provided an individual buffer at the BS. Following the NOMA operation, the BS will send a broadcast signal, $\sum_{i=1}^{V}\sqrt{\alpha_{i}P}s_{i}\left(\tau \right)$, to the destination nodes, where $\alpha_{i}$ is the power allocation coefficient of user $v_i$, $s_{i}\left(\tau \right)$ is the message intended for $v_i$ at time $\left(\tau \right)$, and $P$ is the total  transmit power at the BS. The channel between the BS and the destination nodes are assumed to be block fading, i.e., the fading remains constant during each fading block, but it changes independently from one fading block to another. Meanwhile, the blocklength is assumed to be the same size as of block fading and is taken as $n$. In this work, the channel gains are modeled as  Rayleigh fading distribution with unit variance.  The  users in this  NOMA operation, are classified based on the channel conditions.  The channel coefficent between the user $v_i$ and the BS is referred to by $h_i\left(\tau \right)$, wherein without loss of generality we assume   $\left | h_{1}\left(\tau \right)  \right |^2\leq \left | h_{2}\left(\tau \right)   \right |^2\leq...\leq\left | h_{V}\left(\tau \right)   \right |^2$. Following the NOMA operation, the respective power coefficents are ordered as  $\alpha_{1}\geq\alpha_{2}\geq...\geq \alpha_{V}$. Fig.  \ref{fig:sys2} shows the basic operation of a   NOMA.

The BS  broadcasts a message to  users.  The receive signal at   user  $v_i$   can be formulated as:\footnote{Hereafter, the time index $\tau$ is removed for simplicity, whenever it is clear from context.}
\begin{equation}
\label{eq:recmessage}
y_i=h_i\sum_{i=1}^{V} \sqrt{\alpha_{i}P}s_i + m_i,
\end{equation}
where $y_i$ is the received signal at user $v_i$  and  $m_i$ represents the AWGN.

In  this $V$ users network,   we assume only two-users (out of $V$ users), share the same resource block,   using the NOMA operation. We refer to these users by $v_u$ and $v_t$.  When  $u > t$,  User $v_u$ (strong user)  performs the SIC and detects the  User $v_t$ (weak user)   message. Strong user will then remove the weak  user message from its received message. In this case, received SNR at User  $v_u$  can be formulated as
\begin{equation}
\label{eq:snr_u}
 {\rm{SNR}}_{u}=\alpha_{u}\rho\left | h_u \right |^2 ,
\end{equation}
where $\rho$ is the transmit SNR, i.e., $\rho=\frac{P}{N_{o}B}$, as the $N_{o}B$ is the noise power.  On the other hand, the User $v_u$ message at the weak user will be considered as noise, therefore, User $v_t$ will only decode its own message. The resulting SINR at the weak user is hence given as
\begin{equation}
\label{eq:sinr_t}
{\rm{SINR}}_{t}=\frac{\alpha_{t} \rho \left | h_{t} \right |^2 }{\alpha_{u} \rho \left | h_{t} \right |^2+1}.
\end{equation}
In NOMA operation, the users are  ordered according to their  ordered channel gains. Using  $\rho \left | h_i \right |^2=\gamma_i$, and denoting its  probability density function (PDF) by $f\left ( \gamma_{i} \right )$, the PDF of the ordered $\gamma_i$, $i=\{1,.,.,V\}$,   can be obtained from ordered statistics \cite{R163,R179} and is given as
\begin{align}
\label{eq:pdfclosed_ut}
f_{(i:V)}\left ( \gamma_{i} \right )=\xi_{i}f\left ( \gamma_{i} \right )F\left ( \gamma_{i} \right )^{i-1}\left (1-F\left ( \gamma_{i} \right )       \right )^{V-i},
\end{align}
where $f_{(i:V)}$ is the  PDF of ordered  $\gamma_i$ from a set of V users,    $\xi_{i}=\frac{1}{B\left ( i,V-i+1 \right )}$,  and $B(a,b)$ is the beta function \cite{R182}.  In this work, we consider finite blocklength transmission, hence the achievable rate cannot be represented by the Shannon formula, as proved in \cite{R139}. The results in \cite{R139} show that  the achievable rate not only is  a  function of the received SNR (or SINR), but also the transmission error probability ($\epsilon$) and the transmission blocklength ($n$). Using \cite{R139}, the achievable rate for user $v_u$ and user $v_t$  with finite blocklength can be approximated in bit/s/Hz as,
\begin{align}
&\begin{aligned}
\label{eq:rate_nomas}
r_{u}= {\rm log_2}\left ( 1+\alpha_{u}\gamma_u \right )  - \sqrt{\frac{\delta_{u}}{n}}Q^{-1}{(\epsilon)},
\end{aligned}\\
&\begin{aligned}
\label{eq:rate_nomaw}
r_{t}=  {\rm log_2}\left ( 1+\frac{\alpha_{t}\gamma_t }{\alpha_{u}\gamma_t+1} \right )  -  \sqrt{\frac{\delta_{t}}{n}}Q^{-1}{(\epsilon)},
\end{aligned}
\end{align}
where $\delta_u$  and $\delta_t$  are the channel dispersion for  user  $v_u$ and  user $v_t$   and can be approximated as $\delta_{u}=\sqrt{1-\left ( 1+\alpha_{u}\gamma_u \right )^{-2}}$,  $\delta_{t}=\sqrt{1-\left ( 1+       \frac{\alpha_{t}\gamma_t }{\alpha_{u}\gamma_t+1}  \right )^{-2}}$,  $Q^{-1}(.)$ is the inverse of Gaussian Q-function with  $Q\left ( x \right )=\int_{x}^{\infty}\frac{1}{\sqrt{2\pi}}e^{-\frac{w^2}{2}}dw$, $\epsilon$ is the transmission error probability, and $n$ is the blocklength.
\section{Theory of Effective Capacity}
\label{sec:theory-ec}
In this section, the basic concept for the theory of EC is explained. EC is the dual concept of effective bandwidth and has been proposed in \cite{R147} to introduce the link-layer QoS metrics, such as queuing-delay violation probability   and the probability of non-empty buffer.  Assume  an infinite size buffer   at the BS for   user $v_i$, $i=\{1,.,.,V\}$, and the link capacity (service process) as  $r_{i}(\tau)$ at time $\tau$.  The number of packets arriving at time $\tau$  and the number of packets in queue at $\tau$ are represented by  $a_i(\tau)$  and $q_i(\tau)$,  respectively. Let, the arrival rate and the link-layer capacity be ergodic and stationary process and $E[a_{i}(\tau)]< E[r_{i}(\tau)]$, so the $q_i(s)$ converges to a steady state denoted by $q_i(\infty)$ \cite{R161,R167}. In practice, a buffer overflow will occur if $q_i(\infty)$ exceeds the maximum length of the buffer. Assume, $x$ is a maximum threshold on $q_i(\infty)$, then using  large deviation theorem, we get
 \begin{equation}
\label{eq:ec_theta1}
-\lim\limits_{x \to \infty}\frac{{\rm{ln}}\left ({\rm{Pr}}\left \{q_i(\infty)> x  \right \}  \right )}{x}=\theta_i,
\end{equation}
where $\theta_i$ is the delay exponent of user $v_i$,   $q_i(\infty)$ is the steady state of transmit buffer, and ${\rm{Pr}\{a\textgreater b\}}$ is the probability that  $a \textgreater b$ holds. Now, using EC,  the buffer overflow probability, given in (\ref{eq:ec_theta1}), with a certain target $\theta_i$ can be satisfied, if
\begin{equation}
 \label{eq:ec-theta-lambda}
 \Lambda_{a_i}(\theta_i)+\Lambda_{r_i}(-\theta_i)=0,
\end{equation}
where $\Lambda_{a_i}(\theta_i)=\lim\limits_{T\to\infty} \frac{1}{T} {\rm{log}}(\mathbb{E}[e^{\theta_i \sum_{\tau=1}^{T} a_i(\tau)}])$ is the Garntner-Ellis limits of the source process (arrival rate), and $\Lambda_{r_i}(\theta_i)=\lim\limits_{T\to\infty} \frac{1}{T} {\rm{log}}(\mathbb{E}[e^{\theta_i \sum_{\tau=1}^{T} r_i(\tau)}])$  is the  Garntner-Ellis limits of the service process \cite{R167,R168}. Suppose that, the source rate $a_i(\tau)$ is constant such that $a_i(\tau)=a_i$, now from  (\ref{eq:ec-theta-lambda}) we can get the maximum arrival rate (effective capacity) for some unique $\theta_i$ (delay QoS exponent), which is named effective capacity (EC) and can be approximated by $-\frac{\Lambda_{r_i}(-\theta_i)}{\theta_i}$  \cite{R147}.  From (\ref{eq:ec_theta1}), the delay experienced by the source packets at buffer at time $\tau$ can also be estimated in terms of the queuing delay violation probability  using \cite{R147}
\begin{equation}
 \label{eq:qdvp}
 {\rm{Pr}}\left \{  D_i(\tau)>D_{\rm{max}}^i\right \}\approx {\rm{Pr}}\left \{  q_i(\infty)>0\right \}e^{-\theta_{i}\mu_{i} D_{\rm{max}}^i}.
\end{equation}
The above expression is the queuing delay violation probability  for user $v_i$,  where ${\rm{Pr}}\{q_i(\infty)>0\}$ represents the probability of non-empty buffer, and $D_{\rm{max}}^i$ is the maximum delay. We also note that, $\mu_i=C_{e}^{i}$ \cite{R147}, is the effective capacity satisfying a certain QoS metric for user $v_i$. The value for $\theta_i$ $(\theta_i>0)$ from (\ref{eq:qdvp}) is the decay rate  of the outage probability for the user $v_i$.  A more stringent delay requirements can be represented with a larger value of $\theta_i$, while a smaller value of $\theta_i$ shows a less stringent delay requirement.
\subsection{Effective Capacity in Finite Blocklength Regime}
In this section, we aim to investigate the latency performance of a two-users (out of $V$ users) NOMA  with the short packet communications using the EC framework.  The traditional stochastic model for achievable EC using Shannon limit as the service rate is not suitable when considering the finite blocklength transmission. With short-packet communications in NOMA, we use $r_u$ and $r_t$ as provided in (\ref{eq:rate_nomas}) and (\ref{eq:rate_nomaw}) for the service rate. The stochastic model for the achievable EC with short packet communication is provided in \cite{R154}.  By following the  derived service rate from (\ref{eq:rate_nomas}) and  (\ref{eq:rate_nomaw}),  the EC for the two-users  NOMA  with finite blocklength can be approximated as,
\begin{equation}
\label{eq:1EC}
 C_{e}^{i}=-\frac{1}{\theta_{i}{n}}{\rm{ln}}\left (\mathbb{E} \left [\epsilon+\left ( 1-\epsilon \right )e^{-\theta_{i}{n}{r_i}}  \right ]  \right ),
\end{equation}
where $C_{e}^{i}$ and  $\theta_i$ are the EC and the QoS constraint for  user   $v_i$ respectively, and    $\mathbb{E}[.]$ is the expectation operator.
\section{Effective Capacity of Downlink Two-Users NOMA with Finite Blocklength}
\label{sec:ec-noma-short}
In this section, we derive the achievable EC of a two-users  NOMA and multiple users NOMA network with short packet communications. Focusing on a two-users  NOMA network, we also provide a closed-from expressions for the strong
 and weak users in a finite blocklength regime.
\subsection{Achievable Effective Capacity of Strong-User NOMA with Finite Blocklength}
\noindent Out of the two-users, namely, strong user ($v_u$) and weak user  ($v_t$), first we derive the achievable EC of  the  strong user and then provide its  closed-form expression. Following (\ref{eq:1EC}), the achievable EC of user $v_u$  ($C_{e}^{u}$) is formulated as
\begin{align}
\begin{split}
 \label{eq:1ec_u1}
C_{e}^{u}=-\frac{1}{\theta_{u}{n}}{\rm{ln}}\left (\mathbb{E} \left [\epsilon+\left ( 1-\epsilon \right )
\left ( 1+\alpha_{u}\gamma_u \right )^{2\zeta_{u}} e^{\beta_{u}\delta_{u}}  \right ]  \right ),
\end{split}
\end{align}
where $\zeta_{u}=-\frac{\theta_{u}{n}}{2{\rm{ln}2}}$ and  $\beta_{u}=\theta\sqrt{n}Q^{-1}{(\epsilon)}$.

\noindent The above expression for the achievable EC of a strong user can further be simplified by deriving its closed-form expression. In this regard, after applying the order statistics  from (\ref{eq:pdfclosed_ut}), the achievable EC of a strong user, given in  (\ref{eq:1ec_u1}), is expanded as
\begin{align}
\begin{split}
 \label{eq:1ec_u2}
C_{e}^{u}=-\frac{1}{\theta_{u}{n}}{\rm{ln}}\left (\int_{0}^{\infty}\left (\epsilon+\left ( 1-\epsilon \right )
\left ( 1+\alpha_{u}\gamma_u \right )^{2\zeta_{u}} e^{\beta_{u}\delta_u}  \right )    f_{(u:V)}\left ( \gamma_{u} \right ) d_{\gamma_{u}}  \right ).
\end{split}
\end{align}
The above expression can further be simplified by expanding the order statistics from (\ref{eq:pdfclosed_ut}) and   changing  $e^{\beta_{u}\delta_{u}}$ into a fraction. In this regard, using Maclaurin series for the expansion of $e^{\beta_{u}\delta_{u}}$ such that $e^{\beta_{u}\delta_{u}} \approx 1+  \beta_{u}\delta_{u}  +\frac{\left (\beta_{u}\delta_{u}  \right )^2}{2},$    the achievable EC of a strong user is given by
\begin{align}
 \begin{split}
 \label{eq:1ec_u3}
C_{e}^{u}=&-\frac{1}{\theta_{u}{n}}{\rm{ln}}\Bigg (\int_{0}^{\infty}\left (\epsilon+\left ( 1-\epsilon \right )
\left ( 1+\alpha_{u}\gamma_u \right )^{2\zeta_{u}}  \left (   1+ \beta_{u}\delta_{u} +\frac{\left (\beta_{u}\delta_{u}  \right )^2}{2} \right )   \right )  \\ &   \times \xi_{u}f\left ( \gamma_{u} \right )F\left ( \gamma_{u} \right )^{u-1}\left (1-F\left ( \gamma_{u} \right )       \right )^{V-u} d_{\gamma_{u}}  \Bigg ).
\end{split}
\end{align}
After inserting  $\delta_{u}=\sqrt{1-\left ( 1+\alpha_{u}\gamma_u \right )^{-2}}$  in the above equation, the EC of the strong user can  be formulated as
\begin{align}
 \begin{split}
  \label{eq:1ec_u4}
C_{e}^{u}=&-\frac{1}{\theta_{u}{n}}{\rm{ln}}\left (\int_{0}^{\infty} \left (\epsilon+\left ( 1-\epsilon \right )
\left (  1+\alpha_u\gamma_{u} \right )^{2\zeta_{u}}
+  \beta_{u}   \left (  1+\alpha_u\gamma_{u} \right )^{2\zeta_{u}}   \sqrt{1-\left (  1+\alpha_u\gamma_{u}  \right )^{-2}}    \right. \right. \\&  \left. \left.  +  \frac{{\beta_{u}}^2}{2}   \left (  1+\alpha_u\gamma_{u} \right )^{2\zeta_{u}} \left (  1-\left (  1+\alpha_u\gamma_{u} \right )^{-2} \right )    \right )   \xi_{u}f\left ( \gamma_{u} \right )F\left ( \gamma_{u} \right )^{u-1}\left (1-F\left ( \gamma_{u} \right )       \right )^{V-u} d_{\gamma_{u}}  \right),
\end{split}
\end{align}
now, by inserting  $f\left ( \gamma_{u} \right )=\frac{1}{\rho}e^{-\frac{\gamma_u}{\rho}}$ and $F\left ( \gamma_{u} \right )=1-e^{-\frac{\gamma_u}{\rho}}$ into (\ref{eq:1ec_u4}), we can have
\begin{align}
 \begin{split}
 \label{eq:1ec_u5}
C_{e}^{u}=&-\frac{1}{\theta_{u}{n}}{\rm{ln}}\left (\frac{\xi_{u}}{\rho}\int_{0}^{\infty} \left (\epsilon+\left ( 1-\epsilon \right )
\left (  1+\alpha_u\gamma_{u} \right )^{2\zeta_{u}}
+  \beta_{u}   \left (  1+\alpha_u\gamma_{u}  \right )^{2\zeta_{u}}    \sqrt{1-\left (  1+\alpha_u\gamma_{u}  \right )^{-2}}    \right. \right. \\ &  \left. \left. +    \frac{{\beta_{u}} ^2}{2}   \left (  1+\alpha_u\gamma_{u}  \right )^{2\zeta_{u}} \left ( 1-\left (  1+\alpha_u\gamma_{u} \right )^{-2} \right )   \right )
   e^{-\frac{\left ( V-u+1 \right )\gamma_u}{\rho}}\left (1-e^{-\frac{\gamma_u}{\rho}}  \right )^{u-1} d_{\gamma_{u}}   \right).
\end{split}
\end{align}
After solving the above integral, the final closed-form expression for the achievable EC of the NOMA strong user  in a finite blocklength can be approximated as,
\begin{align}
 \begin{split}
 \label{eq:1ec_u6}
C_{e}^{u}\approx-\frac{1}{\theta_{u}{n}}{\rm{ln}}\Bigg( \epsilon+\left ( 1-\epsilon \right )\Bigg (\frac{\xi_{u}}{\rho\alpha_{u}} \sum_{i=0}^{u-1}\binom{u-1}{i}(-1)^{i}  &  {\rm{H}} \left (1,2+2\zeta_{u},\eta_{u} \right) \left({\rm{K_u+1}} \right)  \\
&  -{\rm{H}} \left(1,2\zeta_{u},\eta_{u}  \right ) \left ({\rm{K_u}}-\frac{\beta_{u}}{2} \right)  \Bigg )   \Bigg ),
\end{split}
\end{align}
where $\eta_{u}=\frac{V-u+1+i}{\rho\alpha_u}$, $K_u=\frac{\beta_{u}^{2}}{2}+\beta_u$, and $\rm{H}(a,b,z)$ is the confluent hypergeometric function of the second kind \cite{R171} such that
\begin{align}
 \begin{split}
 \label{eq:hyper_u}
{\rm{H}}\left ( a,b,z \right )=\frac{1}{\Gamma\left ( a \right )} \int_{0}^{\infty}e^{-zt}t^{a-1}\left ( 1+t \right )^{b-a-1} d_{t} \ \ \rm{for} \ \rm{Re } \ [a],\ \rm{Re} \ [z]>0,
\end{split}
\end{align}
\noindent where $\Gamma\left ( . \right )$ is the Gamma function.
\\
The details for deriving the closed-form expression for strong user are provided in Appendix A. The accuracy of the proposed closed-form expression has also been verified using simulations, as will be shown in Section \ref{sec:numerical-results}.
\subsection{Achievable Effective Capacity of Weak-User NOMA with Finite Blocklength}
Following the steps for deriving the achievable EC of user $v_u$ and its closed-form expression, we can also formulate the achievable EC of user $v_t$ and its closed-form expression.  Using (\ref{eq:1EC}), the achievable EC of a weak user  is given as
\begin{align}
 \begin{split}
 \label{eq:1ec_t1}
C_{e}^{t}=-\frac{1}{\theta_{t}{n}}{\rm{ln}}\left (\mathbb{E} \left [\epsilon+\left ( 1-\epsilon \right )
\left ( \frac{\gamma_t+1 }{\alpha_u\gamma_t+1}  \right )^{2\zeta_{t}} e^{\beta_{t}\delta_{t}}  \right ]  \right ),
\end{split}
\end{align}
where $\zeta_{t}=-\frac{\theta_{t}{n}}{2{\rm{ln}2}}$ and $\beta_{t}=\theta\sqrt{n}Q^{-1}{(\epsilon)}$.

The above expression for the achievable EC of user $v_t$ can be used to derive the closed-form expression. In this sense, applying the order statistics from (\ref{eq:pdfclosed_ut}) in (\ref{eq:1ec_t1}),  we achieve
\begin{align}
 \begin{split}
 \label{eq:1ec_t2}
C_{e}^{t}=-\frac{1}{\theta_{t}{n}}{\rm{ln}}\left (\int_{0}^{\infty} \left (\epsilon+\left ( 1-\epsilon \right )
\left ( \frac{\gamma_t+1 }{\alpha_u\gamma_t+1}  \right )^{2\zeta_{t}} e^{\beta_{t}\delta_{t}}  \right ) f_{(t:V)}\left ( \gamma_{t} \right ) d_{\gamma_{t}}  \right ).
\end{split}
\end{align}
However, the above expression can further be simplified by expanding the order statistics from (\ref{eq:pdfclosed_ut}) and   using the Maclaurin series for the $e^{\beta_{t}\delta_{t}}$ expression such as $e^{\beta_{t}\delta_{t}} \approx 1+  \beta_{t}\delta_{t}  +\frac{\left (\beta_{t}\delta_{t}  \right )^2}{2}$. Now, the achievable EC of user $v_t$ is provided as
\begin{align}
 \begin{split}
 \label{eq:1ec_t3}
C_{e}^{t}=&-\frac{1}{\theta_{t}{n}}{\rm{ln}}\Bigg(\int_{0}^{\infty} \left (\epsilon+\left ( 1-\epsilon \right )
\left ( \frac{\gamma_t+1 }{\alpha_u\gamma_t+1}  \right )^{2\zeta_{t}} \left (   1+ \beta_{t}\delta_{t} +\frac{\left (\beta_{t}\delta_{t}  \right )^2}{2} \right )  \right )  \\ &  \times \xi_{t}f\left ( \gamma_{t} \right )F\left ( \gamma_{t} \right )^{t-1}\left (1-F\left ( \gamma_{t} \right )       \right )^{V-t} d_{\gamma_{t}}  \Bigg),
\end{split}
\end{align}
Using  $\delta_t=\sqrt{1-\left (  \frac{\gamma_t+1 }{\alpha_u\gamma_t+1}   \right )^{-2}}$ in the above expression, the achievable EC of the weak user is  formulated as
\begin{align}
 \begin{split}
 \label{eq:1ec_t4}
C_{e}^{t}=&-\frac{1}{\theta_{t}{n}}{\rm{ln}}\left ( \int_{0}^{\infty} \left (\epsilon+\left ( 1-\epsilon \right )
\left   (  \frac{\gamma_t+1 }{\alpha_u\gamma_t+1}   \right )^{2\zeta_{t}}+  \beta_{t}   \left (  \frac{\gamma_t+1 }{\alpha_u\gamma_t+1}   \right )^{2\zeta_{t}}  \right. \right.  \\ &\times   \sqrt{1-\left (  \frac{\gamma_t+1 }{\alpha_u\gamma_t+1}   \right )^{-2}}    +  \frac{ \beta_{t}   ^2}{2}  \left (  \frac{\gamma_t+1 }{\alpha_u\gamma_t+1}   \right  )^{2\zeta_{t}} \\  & \left. \left.  \times    \left (  1-\left ( \frac{\gamma_t+1 }{\alpha_u\gamma_t+1}   \right )^{-2}  \right )    \right ) \xi_{t}f\left ( \gamma_{t} \right )F\left ( \gamma_{t} \right )^{t-1}\left (1-F\left ( \gamma_{t} \right )       \right )^{V-t} d_{\gamma_{t}} \right).
\end{split}
\end{align}
 Using $f\left ( \gamma_{t} \right )=\frac{1}{\rho}e^{-\frac{\gamma_t}{\rho}}$ and $F\left ( \gamma_{t} \right )=1-e^{-\frac{\gamma_t}{\rho}}$, the above expression for $C_{e}^{t}$ is further simplified to
 \begin{align}
 \begin{split}
 \label{eq:1ec_t5}
C_{e}^{t}=&-\frac{1}{\theta_{t}{n}}{\rm{ln}}\left (\frac{\xi_{t}}{\rho} \int_{0}^{\infty} \left (\epsilon+\left ( 1-\epsilon \right )
\left   (  \frac{\gamma_t+1 }{\alpha_u\gamma_t+1}   \right )^{2\zeta_{t}}+  \beta_{t}   \left (  \frac{\gamma_t+1 }{\alpha_u\gamma_t+1}   \right )^{2\zeta_{t}}  \right. \right.  \\ &\times   \sqrt{1-\left (  \frac{\gamma_t+1 }{\alpha_u\gamma_t+1}   \right )^{-2}}    +  \frac{ \beta_{t}   ^2}{2}  \left (  \frac{\gamma_t+1 }{\alpha_u\gamma_t+1}   \right  )^{2\zeta_{t}} \\  & \left. \left.  \times    \left (  1-\left ( \frac{\gamma_t+1 }{\alpha_u\gamma_t+1}   \right )^{-2}  \right )    \right ) e^{-\frac{\left ( V-t+1 \right )\gamma_t}{\rho}}\left (1-e^{-\frac{\gamma_t}{\rho}}  \right )^{t-1}  d_{\gamma_{t}} \right).
\end{split}
\end{align}
Now, by taking some further mathematical simplification and solving the integrals in (\ref{eq:1ec_t5}), the closed-form expression for the achievable EC of NOMA weak user  in a finite blocklength is
 \begin{align}
 &\begin{aligned}
 \begin{split}
 \label{eq:1ec_t6}
 &C_{e}^{t}\approx -\frac{1}{\theta_{t}n}{\rm{ln}} \left ( \epsilon+\left ( 1-\epsilon \right )    \left ( \Bigg(\frac{\alpha_{u}^{-2\zeta_{t}}\xi_{t} } {\rho} \Bigg( \sum_{r=0}^{t-1}\binom{t-1}{r}(-1)^{r} \frac{1}{\eta_{t}\alpha_{u}}   +\frac{\theta_{t}n(\alpha_{u}-1)}    {\alpha_{u}{\rm{ln}2}} \sum_{r=0}^{t-1}\binom{t-1}{r} \right. \right.   \\ &\times (-1)^{r}e^{\eta_{t}}  {\rm{E_i}} (-\eta_t)+\sum_{s=2}^{\infty}  \binom{2\zeta_{t}}{s}\left (\frac{\alpha_{u}-1}{\alpha_{u}} \right)^s \sum_{r=0}^{t-1}\binom{t-1}{r}(-1)^{r} \Bigg(\frac{\sum_{r=1}^{s-1} \frac{(r-1)!}{\alpha_{u }^{-r}} (-\alpha_u \eta_{t})^{s-r-1}    }{(s-1)!}       \\
& - \frac{(-\alpha_u \eta_{t})^{s-1}}{(s-1)!}        e^{\eta_{t}}    {\rm{E_i}}(-\eta_{t})
  \Bigg)     \Bigg)  \Bigg) \left ( K_t+1 \right)- \left ( \left (  \frac{\alpha_{u}^{-(2\zeta_{t}-2)}\xi_{t} }   {\rho} \left ( \sum_{r=0}^{t-1}  \binom{t-1}{r}(-1)^{r}\frac{1}{\eta_{t}\alpha_{u}}      \right.  \right. \right.  \\
& +\frac{\theta_{t}n(\alpha_{u}-1)}    {\alpha_{u}{\rm{ln}2}}  \sum_{r=0}^{t-1}\binom{t-1}{r} (-1)^{r} e^{\eta_{t}}{\rm{E_i}} (-\eta_t)+\sum_{s=2}^{\infty}   \binom{2\zeta_{t}-2}{s}\left (\frac{\alpha_{u}-1}{\alpha_{u}} \right)^s \sum_{r=0}^{t-1} \binom{t-1}{r}(-1)^{r} \\
&   \left. \left. \left. \left. \left.  \times     \left (\frac{\sum_{r=1}^{s-1} \frac{(r-1)!}{\alpha_{u }^{-r}} (-\alpha_u \eta_{t})^{s-r-1}    }{(s-1)!}-    \frac{(-\alpha_u \eta_{t})^{s-1}}{(s-1)!}        e^{\eta_{t}}    {\rm{E_i}}(-\eta_{t})
  \right )     \right ) \right )  \left ( K_t-\frac{\beta_{t}}{2} \right ) \right ) \right )   \right ),
\end{split}
\end{aligned}
\end{align}
where $\eta_{t}=\frac{V-t+1+r}{\rho\alpha_u}$, $K_t=\frac{\beta_{t}^{2}}{2}+\beta_t$, and ${\rm{E_i}}(.)$ is the exponential integral with  $E_i(x)=-\int_{-x}^{\infty}\frac{e^{-w}}{w}dw$. The detail for deriving the close-form expression for weak user is provided in Appendix B.
\subsection{Achievable Effective Capacity of Multiple NOMA Pairs in Finite Blocklength}
Here, we investigate the total achievable EC of multiple NOMA pairs with finite blocklength. Specifically, we consider that, there are in total  $V$ users, which are divided into $\frac{V}{2}$ pairs, such that $\mathbb{M}=\left \{ 1,2,...,\frac{V}{2} \right \}$ which shows the group index. We introduce  $\phi$ as the combination of all NOMA pairs such that $\phi=\left \{\phi_1,\phi_2,...,\phi_{V/2}  \right \}$. We take the $m^{th}$ NOMA pair with finite blocklength with the  strong user denoted as $v_{u_m}$ and  weak user as  $v_{t_m}$   such that  $\phi_{m}=\left \{ ({t_m},u_m) \ | \  {t_m}\neq u_m, \ \left | h_{t_m} \right |^2\leq \left | h_{u_m} \right |^2,\ \forall_{m} \in \mathbb{M} \right \}$. We will consider in our work, the $m^{th}$ NOMA pair, and the achievable EC for the $v_{u_m}$ and  $v_{t_m}$  user will be investigated. However, for the inter-pair multiple access time-division multiple access (TDMA) will be used. As we are considering the $m^{th}$ NOMA pair with finite blocklength regime, the transmission rate for the strong and  weak users  can be approximated as
\begin{align}
&\begin{aligned}
 \begin{split}
 \label{eq:lemma2_rate_u}
r_{u_m}=\frac{2}{V}\left({\rm{log_2}}\left (1+ \alpha_{u_m}\gamma_{u_m}   \right )- \sqrt{\frac{\delta_{u_m}}{n}}Q^{-1}{(\epsilon)}  \right ),
\end{split}
\end{aligned}\\
&\begin{aligned}
 \begin{split}
\label{eq:lemma2_rate_t}
r_{t_m}=\frac{2}{V}\left({\rm{log_2}}\left (1+ \frac{\gamma_{t_m}+1 }{\alpha_{u_m}\gamma_{t_m}+1}   \right )- \sqrt{\frac{\delta_{t_m}}{n}}Q^{-1}{(\epsilon)}  \right ),
\end{split}
\end{aligned}
\end{align}
where $\gamma_{i_m}=\rho\left | h{i_m} \right |^{2}$. Using   (\ref{eq:lemma2_rate_u}) and (\ref{eq:lemma2_rate_t})   as the transmission rate with finite blocklength and applying the Gartner-Ellis theorem, the achievable EC for the   strong user and weak  user can be formulated as,
 \begin{align}
 &\begin{aligned}
 \begin{split}
\label{eq:lemma2_ec_u}
 C_{e}^{u_m}=-\frac{1}{\theta_{u_m}{n}}{\rm{ln}}\left (\mathbb{E} \left [\epsilon+\left ( 1-\epsilon \right )
\left ( 1+   \alpha_{u_m}\gamma_{u_m} \right )^{\frac{4\zeta_{u_m}}{V}}   e^{\frac{2}{V}\beta_{u_m}\delta_{u_m}}  \right ]  \right ),
\end{split}
\end{aligned}\\
&\begin{aligned}
 \begin{split}
\label{eq:lemma2_ec_t}
C_{e}^{t_m}=-\frac{1}{\theta_{t_m}{n}}{\rm{ln}}\left (\mathbb{E} \left [\epsilon+\left ( 1-\epsilon \right )
\left ( 1+     \frac{\gamma_{t_m}+1 }{\alpha_{u_m}\gamma_{t_m}+1}  \right ) ^{\frac{4\zeta_{t_m}}{V}}    e^{\frac{2}{V}\beta_{t_m}\delta_{t_m}}  \right ]  \right ).
\end{split}
\end{aligned}
\end{align}
The achievable EC of multiple NOMA pairs in (\ref{eq:lemma2_ec_u}) and (\ref{eq:lemma2_ec_t}), and the EC of two users NOMA network in (\ref{eq:1ec_u1}) and (\ref{eq:1ec_t1}), have  similar expressions Therefore, by following Appendix A and Appendix B, the closed-form expressions for the achievable EC of  $v_{t_m}$ and $v_{u_m}$ users in multiple NOMA pairs with finite blocklength can be derived.  Total EC ($T_{ec}$) can be estimated by using $\sum_{m=1}^{V} (C_{e}^{t_m}+C_{e}^{u_m})$. Analytical results in Section \ref{sec:numerical-results} regarding the multiple NOMA pairing setting have been investigated in detail. The users with  more distinct and less distinct channel conditions have been paired together and their $T_{\rm{ec}}$ has been analyzed with respect to $\rho$.
\section{Effective Capacity of downlink two-users   NOMA with Finite Blocklength At High Transmit SNRs}
The performance of a two-users NOMA in a finite blocklength regime can be investigated  by taking into consideration the closed-form expressions for strong user and weak user's EC presented in (\ref{eq:1ec_u6}) and (\ref{eq:1ec_t6}), respectively. However, the obtained closed-form expressions are complicated and complex to understand.  Here, we provide an approximation to simplify the achievable EC formulation for the strong and weak user NOMA network and their closed-form expressions. In this regard, the channel dispersion $(\delta_i)$ is approximated as $ \delta_i \approx1$,  for high SNR \cite{R139}. Considering this above approximation at high SNR, contrary to  (\ref{eq:1ec_u1}) and (\ref{eq:1ec_t1}), the  achievable EC of strong and weak user can now be given as
\begin{align}
&\begin{aligned}
\begin{split}
 \label{eq:2ec_ub1}
\bar{C}_{e}^{u}=-\frac{1}{\theta_{u}{n}}{\rm{ln}}\left (\mathbb{E} \left [\epsilon+\left ( 1-\epsilon \right )
\left ( 1+\alpha_{u}\gamma_u \right )^{2\zeta_{u}} e^{\beta_{u}}  \right ]  \right ),
\end{split}
\end{aligned}\\
&\begin{aligned}
 \begin{split}
 \label{eq:2ec_tb1}
\bar{C}_{e}^{t}=-\frac{1}{\theta_{t}{n}}{\rm{ln}}\left (\mathbb{E} \left [\epsilon+\left ( 1-\epsilon \right )
\left ( \frac{\gamma_t+1 }{\alpha_u\gamma_t+1}  \right )^{2\zeta_{t}} e^{\beta_{t}}  \right ]  \right ),
\end{split}
\end{aligned}
\end{align}
where $\bar{C}_{e}^{u}$ and $\bar{C}_{e}^{t}$ are the achievable EC of strong user and weak user NOMA at high transmit SNR ($\delta_i=1$), respectively. The above expressions can be further simplified by deriving their closed-form expressions. Using the order statistics from (\ref{eq:pdfclosed_ut}), the achievable EC of strong and weak user at high transmit SNR can be expanded as
\begin{align}
&\begin{aligned}
\begin{split}
 \label{eq:2ec_ub2}
\bar{C}_{e}^{u}=-\frac{1}{\theta_{u}{n}}{\rm{ln}}\left (\int_{0}^{\infty}\left (\epsilon+\left ( 1-\epsilon \right )
\left ( 1+\alpha_{u}\gamma_u \right )^{2\zeta_{u}} e^{\beta_{u}}  \right ) f_{(u:V)}\left ( \gamma_{u} \right ) d_{\gamma_{u}}  \right ),
\end{split}
\end{aligned}\\
&\begin{aligned}
 \begin{split}
 \label{eq:2ec_tb2}
\bar{C}_{e}^{t}=-\frac{1}{\theta_{t}{n}}{\rm{ln}}\left (\int_{0}^{\infty} \left (\epsilon+\left ( 1-\epsilon \right )
\left ( \frac{\gamma_t+1 }{\alpha_u\gamma_t+1}  \right )^{2\zeta_{t}} e^{\beta_{t}}  \right ) f_{(t:V)}\left ( \gamma_{t} \right ) d_{\gamma_{t}}  \right ).
\end{split}
\end{aligned}
\end{align}
The above integrals can be solved using similar steps as in Appendix A and Appendix B. After solving the above integrals, the final closed-form expressions for the achievable EC of strong and weak users at high SNR can be approximated as
\begin{align}
&\begin{aligned}
 \begin{split}
 \label{eq:2ec_ub3}
\bar{C}_{e}^{u}\approx-\frac{1}{\theta_{u}{n}}{\rm{ln}}\left ( \epsilon+\left ( 1-\epsilon \right )\left (\frac{\xi_{u}}{\rho\alpha_{u}} e^{\beta_{u}} \sum_{i=0}^{u-1}\binom{u-1}{i}(-1)^{i}  {\rm{H}} \left (1,2+2\zeta_{u},\eta_{u} \right)   \right )   \right ),
\end{split}
\end{aligned}\\
&\begin{aligned}
 \begin{split}
 \label{eq:2ec_tb3}
&\bar{C}_{e}^{t}\approx-\frac{1}{\theta_{t}n}{\rm{ln}} \left ( \epsilon+\left ( 1-\epsilon \right )    \frac{\alpha_{u}^{-2\zeta_{t}}\xi_{t} } {\rho} e^{\beta_{t}} \Bigg( \sum_{r=0}^{t-1}\binom{t-1}{r}(-1)^{r} \frac{1}{\eta_{t}\alpha_{u}}   +\frac{\theta_{t}n(\alpha_{u}-1)}    {\alpha_{u}{\rm{ln}2}} \sum_{r=0}^{t-1}\binom{t-1}{r} \right. \\ &  \times (-1)^{r} e^{\eta_{t}} {\rm{E_i}} (-\eta_t)+\sum_{s=2}^{\infty}  \binom{2\zeta_{t}}{s}\left (\frac{\alpha_{u}-1}{\alpha_{u}} \right)^s \sum_{r=0}^{t-1}\binom{t-1}{r}(-1)^{r} \Bigg(\frac{\sum_{r=1}^{s-1} \frac{(r-1)!}{\alpha_{u }^{-r}} (-\alpha_u \eta_{t})^{s-r-1}    }{(s-1)!}  \\ &  \left. - \frac{(-\alpha_u \eta_{t})^{s-1}}{(s-1)!}          e^{\eta_{t}}    {\rm{E_i}}(-\eta_{t})
  \Bigg)     \Bigg)    \right ).
\end{split}
\end{aligned}
\end{align}
%
%
\subsection{Effective Capacity of Downlink Two-Users NOMA with Finite Blocklength at Extremely High Transmit SNR ($\rho\to \infty$)}
We also investigate the impact of the extremely high transmit SNR on the achievable EC of two users  NOMA network. In this regard,  the achievable EC of the NOMA strong and weak users at extremely high transmit SNR can be derived by inserting the   $\rho \rightarrow \infty$ in the EC formulation.  Using the (\ref{eq:1ec_u1}) and  inserting the $\rho \rightarrow \infty$,  the EC of the strong user at extremely high SNR can be expressed as,
\begin{align}
\begin{split}
\label{eq:app_c1}
&{\rm{lim_{\rho \rightarrow \infty}}}-\frac{1}{\theta_{u}{n}}{\rm{ln}}\left (\mathbb{E} \left [ \epsilon+\left ( 1-\epsilon \right )   e^{-\theta_{u}{n}{({\rm{log_2}}( 1+\alpha_{u}\gamma_u )-\sqrt{\frac{1-\left (  1+\alpha_{u}\gamma_u \right )^{-2}}{n}}Q^{-1}{(\epsilon)})}}  \right ]  \right )\\
&=-\frac{1}{\theta_{u}n}\rm{log}(\epsilon).
\end{split}
\end{align}
Likewise, using the (\ref{eq:1ec_t1}) and inserting the $\rho \rightarrow \infty$, the EC of the NOMA weak user at extremely high transmit SNR can be expressed as,
\begin{align}
\begin{split}
\label{eq:app_c2}
&\lim_{\rho \rightarrow \infty}-\frac{1}{\theta_{t}{n}}{\rm{ln}}\left (\mathbb{E} \left [\epsilon+\left ( 1-\epsilon \right )
\left ( \frac{\gamma_{t}+1 }{\alpha_{u}\gamma_t+1}  \right )^{2\zeta_{t}}  e^{\beta_{t}\sqrt{1-\left ( \frac{\gamma_{t}+1 }{\alpha_{u}\gamma_t+1}  \right )^{-2}}}  \right ]  \right )\\
&=-\frac{1}{\theta_{t}{n}}{\rm{ln}}\left ( \epsilon+\left ( 1-\epsilon \right )  {\alpha_{u}^{-2\zeta_{t}}} e^{\beta_{t}\sqrt{1-\alpha_u^{2}}}    \right ).
\end{split}
\end{align}
It is clear that, when the  the transmit $\rho$ is extremely high, then EC is limited  (upper bounded) by a fixed value which is not a function of $\rho$.  The achievable EC of the strong user is limited by delay exponent, transmission error penalty, and blocklength. In the case of weak user, it is clear from the (\ref{eq:app_c2}) that the achievable EC is limited by $-\frac{1}{\theta_{t}{n}}{\rm{ln}}\left ( \epsilon+\left ( 1-\epsilon \right )  {\alpha_{u}^{-2\zeta_{t}}} e^{\beta_{t}\sqrt{1-\alpha_u^{2}}}    \right )$.  When the transmit $\rho$ is extremely high, the achievable EC of weak user is limited by transmission penalty due to short packet communications, transmission error probability, and the power co-efficient.
\section{Numerical Results}
\label{sec:numerical-results}
In this section, we  evaluate, through numerical simulations, the performance of the  proposed
NOMA network with finite blocklength. We will also confirm the accuracy of the proposed closed-form expressions and  lemma   that are proposed in Section \ref{sec:ec-noma-short}. The total number of users  is taken as $V=10$, and the  $2^{nd}$ and $8^{th}$ users are paired together. These are users with $2^{nd}$ and $8^{th}$   weakest channels such that, $t=2$, and $u=8$. The respective power coefficients of the users are taken as $\alpha_t=0.8$ and $\alpha_u=0.2$, and  ($n=400$), unless otherwise specified.

\begin{figure}
\centering
  \includegraphics [width=0.8\linewidth]{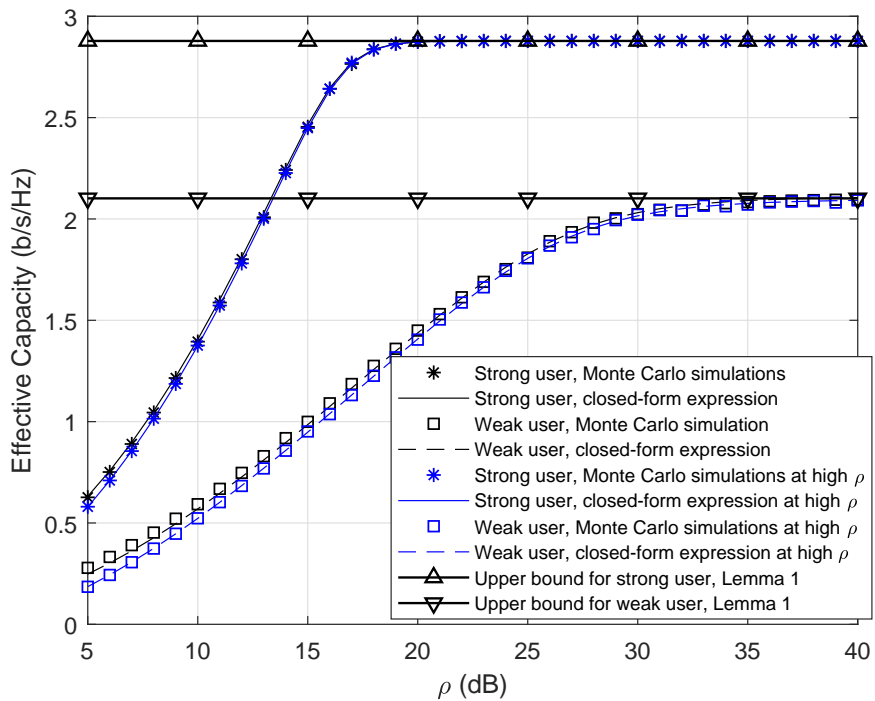}
\caption{Effective Capacity of NOMA weak user and strong user versus transmit SNR in dB, $\theta=0.01$, $n=400$, and $\epsilon=10^{-5}$.}
\label{fig:closed-monte}
\end{figure}
\begin{figure}
\centering
  \includegraphics[width=0.8\linewidth]{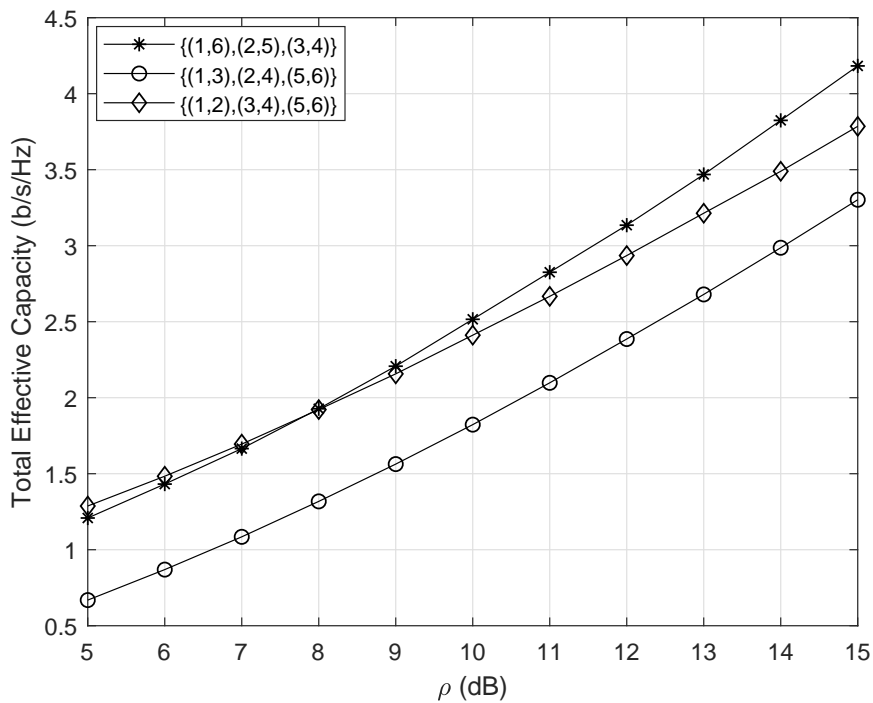}
\caption{Total achievable EC of multiple NOMA pairs versus transmit SNR in dB, $\theta=0.01$,   $\epsilon=10^{-5}$, and $V=6$. }
\label{fig:lemma2}
\end{figure}
\begin{figure}
\centering
  \includegraphics[width=0.8\linewidth]{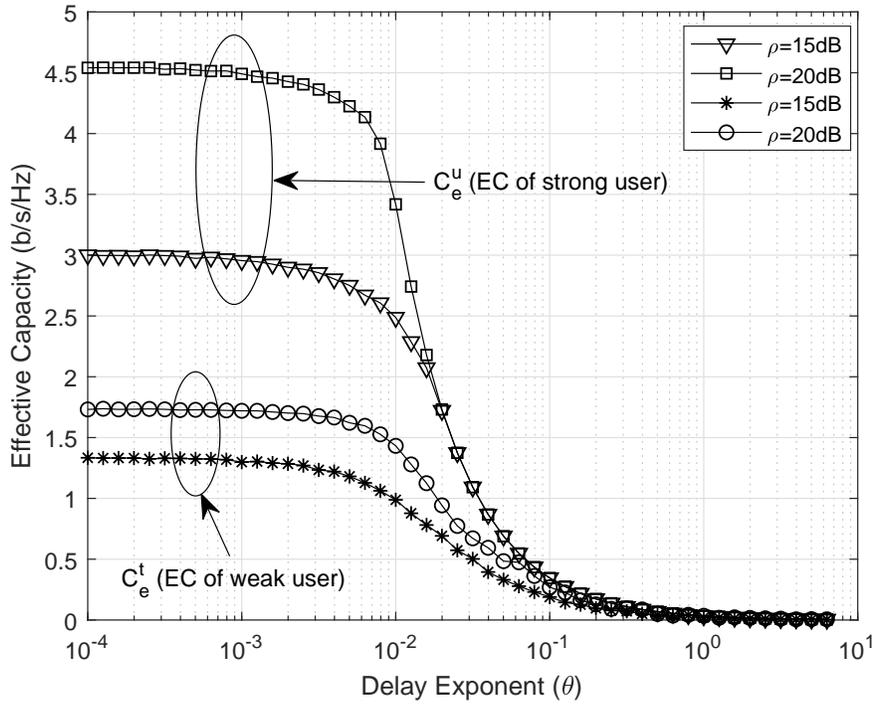}
\caption{Effective Capacity of NOMA strong and weak user versus delay exponent $\theta$ and  $(\epsilon)=10^{-6}$.    }
\label{fig:ec-theta-weak}
\end{figure}
\begin{figure}
\centering
  \includegraphics[width=0.8\linewidth]{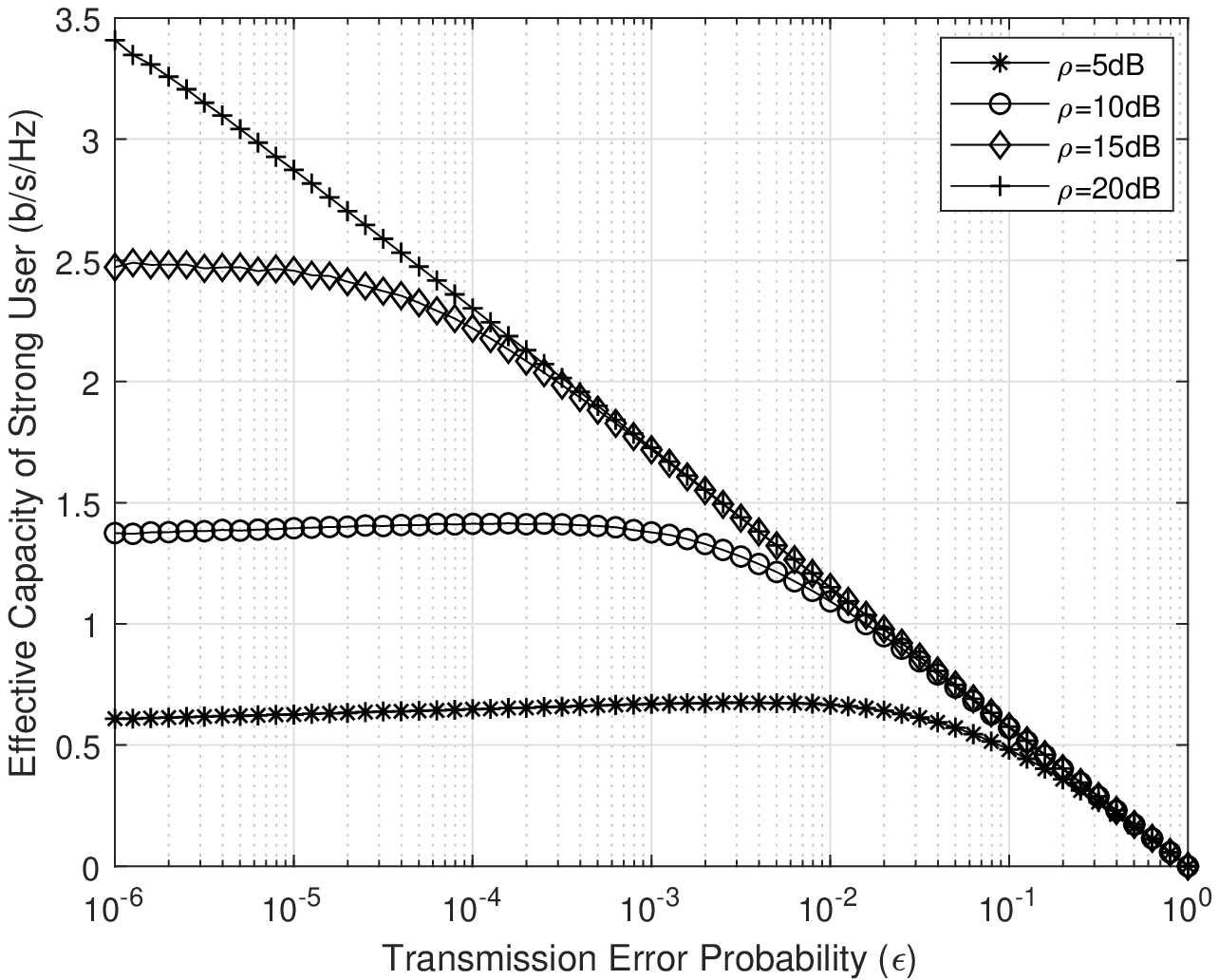}
\caption{Effective Capacity of NOMA strong user versus transmission error probability $(\epsilon)$ with $n=400$ and  $\theta=0.01$. }
\label{fig:ec-tr-strong}
\end{figure}
\begin{figure}
\centering
  \includegraphics[width=0.8\linewidth]{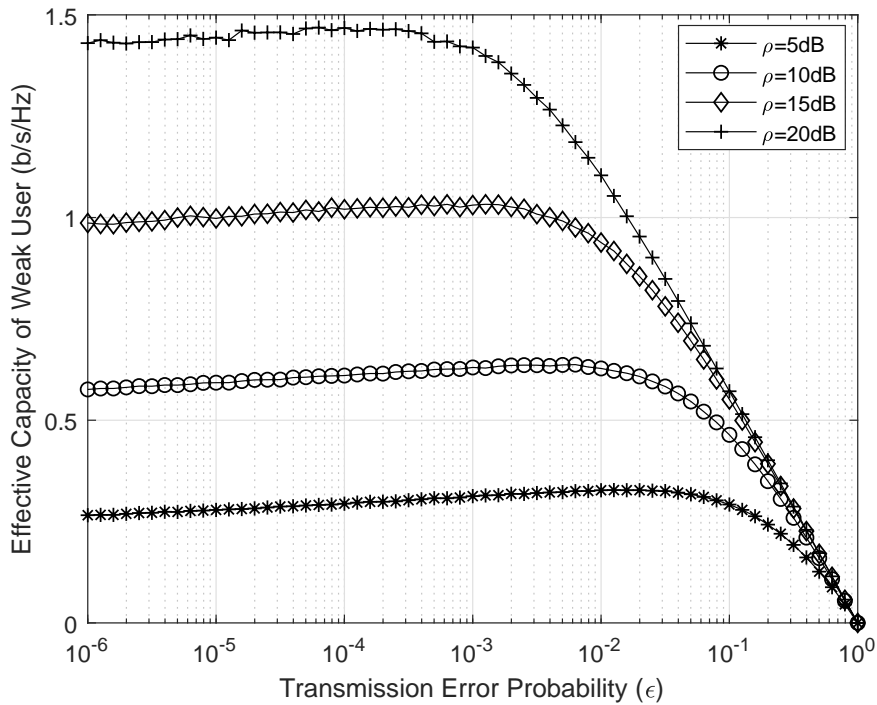}
\caption{Effective Capacity of NOMA weak user versus transmission error probability $(\epsilon)$ with $n=400$ and  $\theta=0.01$. }
\label{fig:ec-tr-weak}
\end{figure}

The accuracy of the closed-form expressions for the EC of the  NOMA users with finite blocklength are investigated in Fig. \ref{fig:closed-monte}. The figure  shows the plots of the $C_{e}^{u}$ (strong-user)  and $C_{e}^{t}$ (weak user)  in b/s/Hz  versus transmit SNR ($\rho$) in dB, where  $\epsilon=10^{-5}$, and the delay exponent $\theta=0.01$. The results for these curves have been obtained using the  proposed extended and the simple  closed-form expressions, i.e,  (\ref{eq:1ec_u6}), (\ref{eq:2ec_ub3}), (\ref{eq:1ec_t6}), and (\ref{eq:2ec_tb3})  and the  Monte Carlo simulations. The accuracy of the closed-form expression for the strong user and the weak user can be confirmed. However,  a small mismatch  exists between the simulations when     using the approximation of $e^{\beta_{i}\delta_{i}} \approx 1+  \beta_{i}\delta_{i}  +\frac{\left (\beta_{i}\delta_{i}  \right )^2}{2}$ (for deriving the extended closed-form expression given in  (\ref{eq:1ec_u6}), and (\ref{eq:1ec_t6})  ) and  using  $\delta_i\approx 1$ at high SNR (for deriving simple closed-form expression given in (\ref{eq:2ec_ub3}) and (\ref{eq:2ec_tb3})). Fig. 1 also confirms the accuracy of the Lemma 1. The achievable EC of strong and weak user is upper bounded when the transmit SNR becomes high. However, the achievable EC of strong user and weak user saturates at different values of $\rho$.

Fig. \ref{fig:lemma2} shows the plots of $T_{\rm{ec}}$ versus transmit SNR for different user pairing set $\phi$ of multiple NOMA pairs network with finite blocklength. The delay exponent for all  users is $\theta=0.01$, $n=400$,  $\epsilon=10^{-5}$,   $V=6$, and  $\forall_m=\mathbb{M}$. Various set of users,  depending on their channel conditions, have been paired together. Impact of transmit SNR on the $T_{ec}$ of different pairing sets are simulated in Fig. \ref{fig:lemma2} which shows that the paring set  $\phi=\left \{ \left ( 1,6 \right ),\left ( 2,5 \right ),\left ( 3,4 \right ) \right \}$ provides the higher $T_{ec}$ as compared to the other pairing set. This shows  that the users with distinct channel conditions when paired together achieve higher $T_{ec}$ as compared to the users with less distinct channel conditions.

\begin{figure}
\centering
  \includegraphics[width=0.8\linewidth]{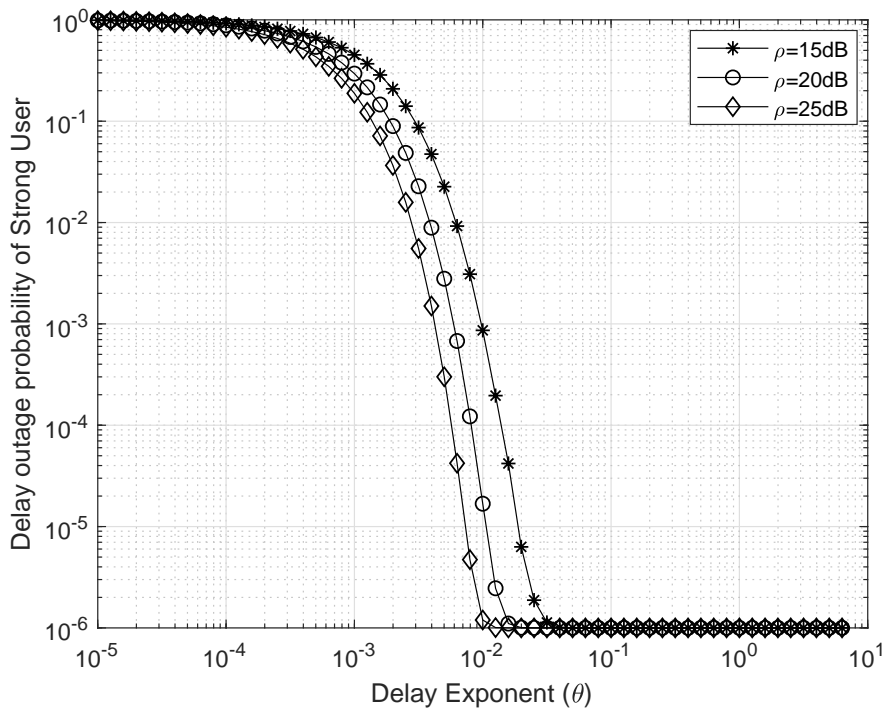}
\caption{Queuing Delay Violation probability versus QoS exponent   constraint ($\theta$) for strong user , $D_{\rm{max}}=400$,  $\epsilon=10^{-6}$, and $n=400$.}
\label{fig:qdvp-theta-strong}
\end{figure}
\begin{figure}
\centering
  \includegraphics[width=0.8\linewidth]{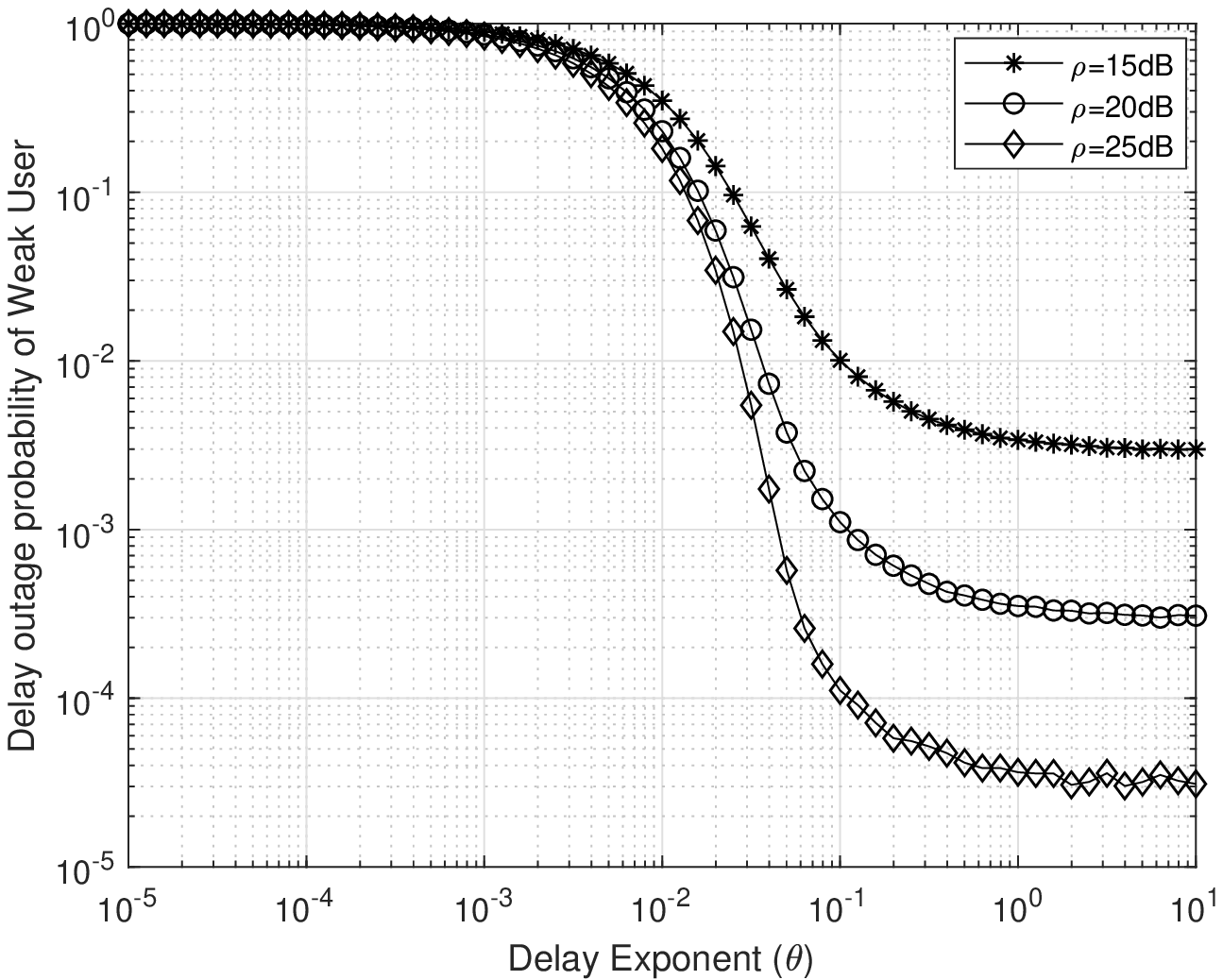}
\caption{Queuing Delay Violation probability versus QoS exponent   constraint ($\theta$) for weak user , $D_{\rm{max}}=400$,  $\epsilon=10^{-6}$, and $n=400$.}
\label{fig:qdvp-theta-weak}
\end{figure}
\begin{figure}
\centering
  \includegraphics[width=0.8\linewidth]{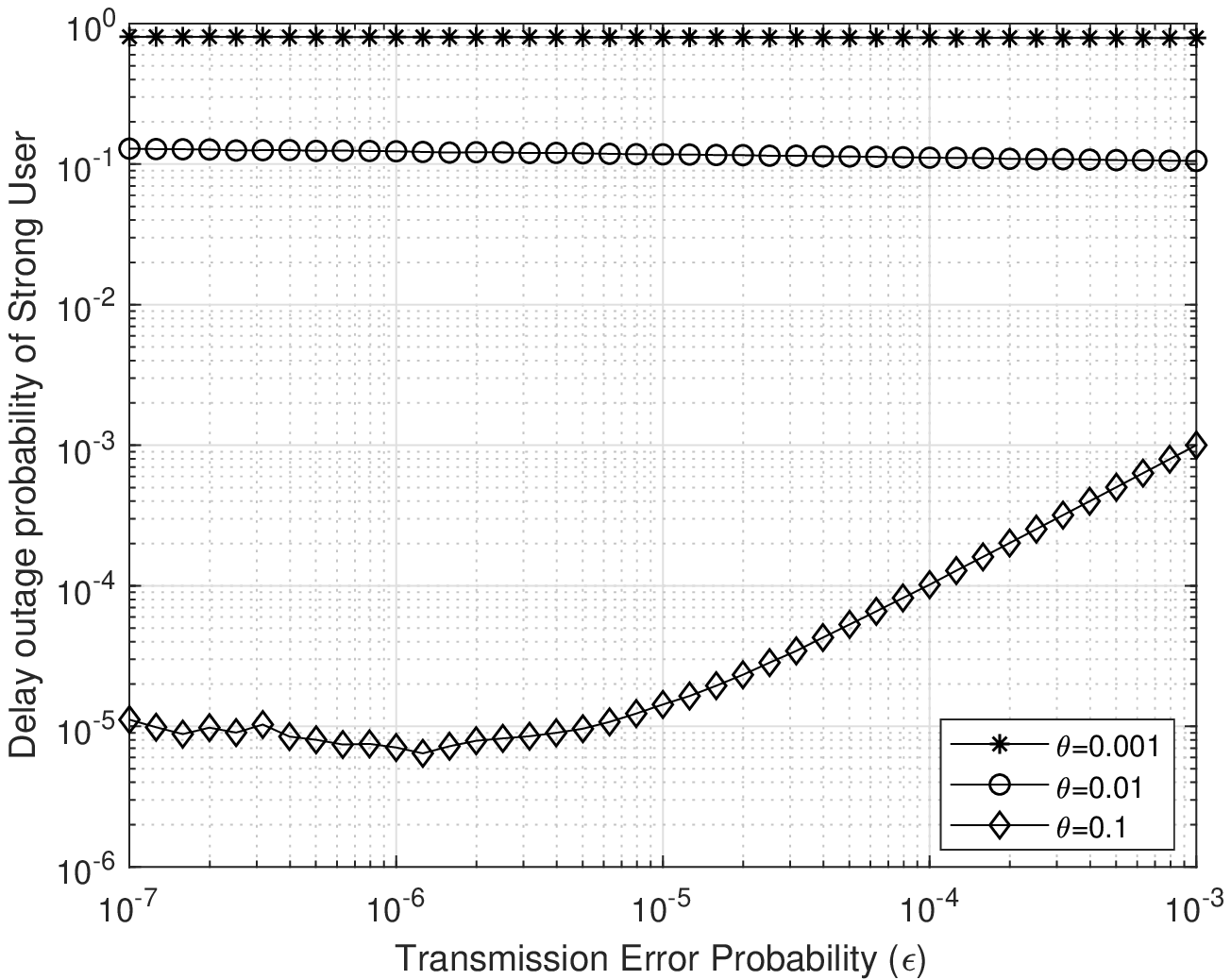}
\caption{Queuing Delay Violation probability versus transmission error probability ($\epsilon$) for strong user , $D_{\rm{max}}=100$,   $n=100$, and $\rho=20$dB.}
\label{fig:qdvp-tr-strong}
\end{figure}
\begin{figure}
\centering
  \includegraphics[width=0.8\linewidth]{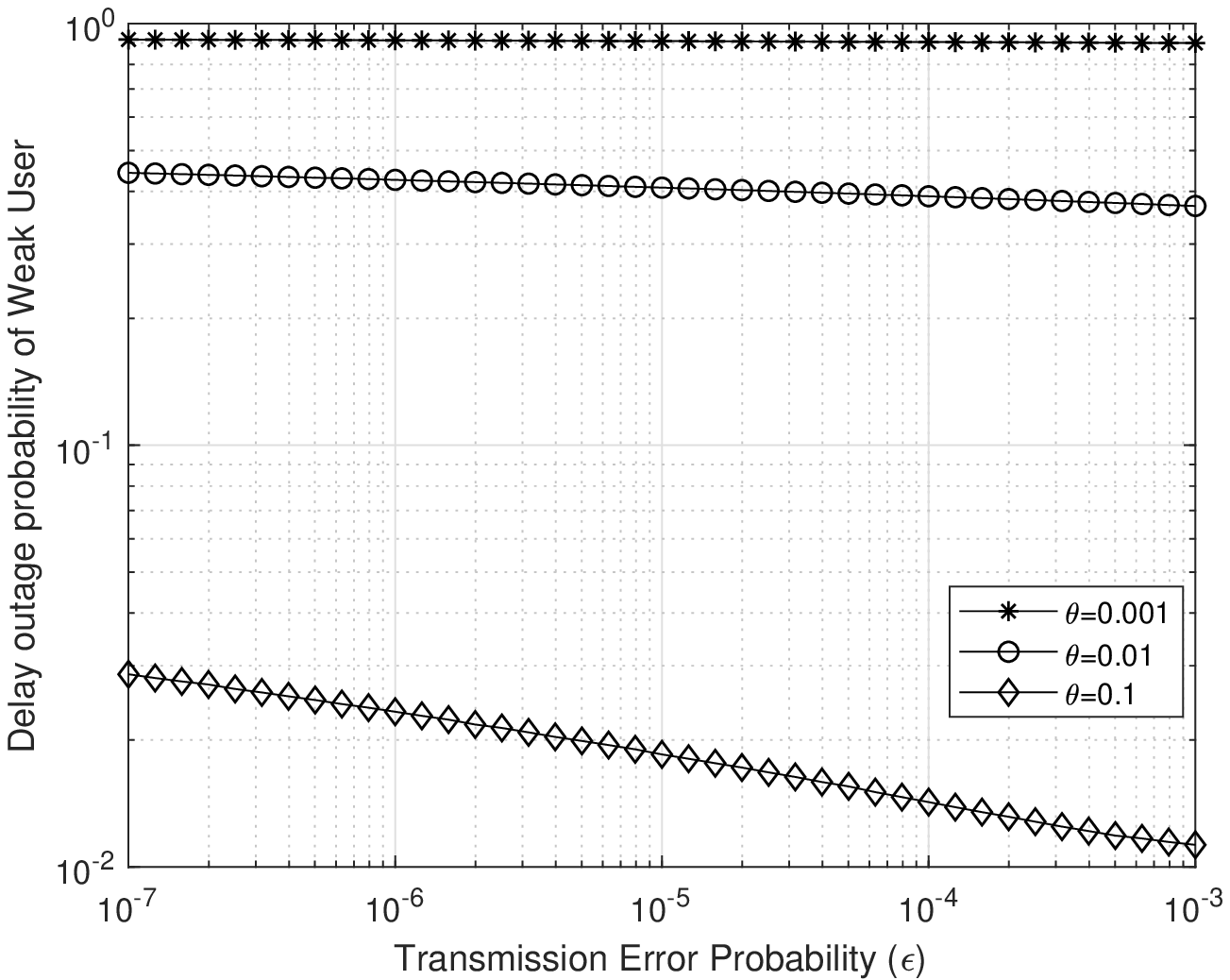}
\caption{Queuing Delay Violation probability versus transmission error probability ($\epsilon$) for weak user , $D_{\rm{max}}=100$,   $n=100$, and $\rho=20$dB.}
\label{fig:qdvp-tr-weak}
\end{figure}

Fig. \ref{fig:ec-theta-weak} shows the plots of the achievable EC of strong user and weak user versus delay exponent $\theta$ with $\rho=[15\rm{dB},20\rm{dB}]$, $n=400$, and $\epsilon=10^{-6}$. It is clear that, the achievable EC of both users decreases when the delay exponent becomes stringent. More specifically, the gain in EC of strong user at the loose delay requirements (low values of $\theta$) is more significant (with large gap) at the same values of $\rho$ as compared to the weak user. However, as the delay exponent becomes stringent, the EC of weak user seems to be more stable as compared to the strong user, i.e., weak user can tolerate more stringent delay.

One of the finite blocklength features, i.e., transmission error probability $\epsilon$ for the achievable EC of the two-user  NOMA has been analysed in Fig. \ref{fig:ec-tr-strong} and Fig. \ref{fig:ec-tr-weak}.  In Fig. \ref{fig:ec-tr-strong}, the EC of strong user is plotted versus  $\epsilon$ for various values of $\rho$, while the blocklength is set to $n=400$ and $\theta=0.01$. We refer the readers to  (\ref{eq:1ec_u1}) for further clarifying the behavior of the plots of this figure.  It is clear that,  when the $\rho$ is very small, i.e., for 5dB,  the term $\Big( \left ( 1-\epsilon \right ) ( 1+\alpha_{u}\gamma_u  )^{2\zeta_{u}} e^{\beta_{u}\delta_{u}} \Big)$  from the EC formulation in (\ref{eq:1ec_u1})  is big as compared to the $\epsilon$. This results into sudden decrease in the achievable EC at low $\rho$ and higher values of transmission error probability. However, as the value of $\rho$ increases, the $\epsilon$ factor becomes dominant, that results into the minimum EC gain yet more stable as compared to the low $\rho$.

Fig. \ref{fig:ec-tr-weak} plots the curves of the EC of weak user NOMA versus $\epsilon$ with various values of $\rho$. As compared to the Fig. \ref{fig:ec-tr-strong}, this shows a considerable decrease in the EC due to the weak channel conditions of the user $v_t$. We refer the readers to  (\ref{eq:1ec_t1}) for further clarification on the behavior of the plots of this figure. As compared to the strong user, the $\epsilon$ factor (due to short packet communication) in weak user achievable EC  remains more dominant as compared to the $\Big(\left ( 1-\epsilon \right ) \left ( \frac{\gamma_t+1 }{\alpha_u\gamma_t+1}  \right )^{2\zeta_{t}} e^{\beta_{t}\delta_{t}}\Big)$ factor, even at the higher values (20dB) of  $\rho$. However, when the value of  $\epsilon$ in simulation increases, the steady trend in EC diminishes, and at the very high values of  $\epsilon$, the EC of weak user becomes zero.

Fig. \ref{fig:qdvp-theta-strong} includes the curves of queuing delay violation probability and the delay exponent $(\theta)$. The delay threshold is set to  $D_{\rm{max}}=400$,  $n=400$, and $\epsilon=10^{-6}$. The trend of plots in this figure can well be understood by following the EC formulation of strong user from  (\ref{eq:1ec_u1}).  It is clear that, as the $\theta$ becomes more stringent, the queueing delay violation probability cannot be improved further below a certain value. This is due to the dominance of $\epsilon$  as compared to the  $\Big( \left ( 1-\epsilon \right ) ( 1+\alpha_{u}\gamma_u )^{2\zeta_{u}} e^{\beta_{u}\delta_{u}} \Big)$ term from EC formulation.  At the high value of $\theta$,  $\Big( \left ( 1-\epsilon \right ) ( 1+\alpha_{u}\gamma_u )^{2\zeta_{u}} e^{\beta_{u}\delta_{u}} \Big)$ factor is very small, and hence, $\epsilon$ becomes the dominant factor.  However,  strong user shows a considerably high improvement in queuing delay violation probability as compared to the weak user, this is due to the better channel conditions.

Fig. \ref{fig:qdvp-theta-weak} shows the plots of queuing delay violation probability versus delay exponent for various values of $\rho$ for the weak user.  As here, the SINR and weak channel conditions are in focus, these result into the queuing delay violation probability to be restricted at different values versus $\theta$.  As $\theta$ becomes  stringent,  queuing delay violation probability does not improve below a certain limit due to the characteristics of short packet communication. However, when compared to the strong user, weak user does not show a considerable improvement in queueing delay violation probability.

In Fig. \ref{fig:qdvp-tr-strong} and Fig. \ref{fig:qdvp-tr-weak}, interesting trends of queuing delay violation probability and $\epsilon$ for various values of delay exponent $\theta$ have been analysed for the strong  and weak users with $D_{\rm{max}}=100$, $\rho=20$, and blocklength $n=100$. It is noted that, when the delay requirements are loose, $\epsilon$ does not have any significant impact on the queuing delay violation probability.   However, when the delay exponent becomes more stringent, $\epsilon$ has a significant impact on the queuing delay violation probability. It further confirms that, when the delay exponent becomes stringent,  queuing delay violation probability does not improve below a certain value to the dominance factor of $\epsilon$.

Fig. \ref{fig:qdvp-tr-weak} shows the curves of the plots for queuing delay violation probability versus $\epsilon$ for different values of delay exponent for the user with weak channel condition. It is evident that  the impact of queuing delay violation probability on ${\epsilon}$ is not very significant when the delay exponent is loose. However, when the delay exponent is stringent  $\theta \rightarrow [0.1]$,  $\epsilon$ has a significant impact on the queuing delay violation probability. However,   as compared to the strong user, the weak user does not show much improvement in  queuing delay violation probability. This result also confirms the impact of short packet communication on the queuing delay violation probability of weak user.
\section{Conclusion}
\label{sec:conclusion}
Effective capacity (EC)-based performance analysis of a two-users (out of $V$ users)  non-orthogonal multiple access (NOMA) network in finite blocklength regime  was investigated in detail in this paper. Overall reliability requirements   was analysed by taking into consideration the queuing delay violation probability and the transmission error probability. We derived the closed-form expressions for the individual EC of a strong and weak users and confirmed their accuracy using the Monte-Carlo simulation. We also investigated the achievable EC of a  multiple pair NOMA pairs in finite blocklength regime, and showed that users with distinct channel conditions achieves more EC as compared to the users with less distinct channel conditions.  We found that for two user  NOMA, loose delay requirements do not have a significant impact on the queuing delay violation probability under  transmission error probability constraint. However, when the delay exponent becomes more stringent, the queueing delay violation probability could not be improved below a certain value under transmission error probability constraint.
%
%
%
%
\section*{APPENDIX A}
The achievable EC with order statistics from (\ref{eq:1ec_u5}) is given as,
\begin{align}
 \begin{split}
 \label{eq:app_1ec_u5}
C_{e}^{u}=&-\frac{1}{\theta_{u}{n}}{\rm{ln}}\left (\frac{\xi_{u}}{\rho}\int_{0}^{\infty} \left (\epsilon+\left ( 1-\epsilon \right )
\left (  1+\alpha_u\gamma_{u} \right )^{2\zeta_{u}}
+  \beta_{u}   \left (  1+\alpha_u\gamma_{u}  \right )^{2\zeta_{u}}    \sqrt{1-\left (  1+\alpha_u\gamma_{u}  \right )^{-2}}    \right. \right. \\ &  \left. \left. +    \frac{{\beta_{u}} ^2}{2}   \left (  1+\alpha_u\gamma_{u}  \right )^{2\zeta_{u}} \left ( 1-\left (  1+\alpha_u\gamma_{u} \right )^{-2} \right )   \right )
   e^{-\frac{\left ( V-u+1 \right )\gamma_u}{\rho}}\left (1-e^{-\frac{\gamma_u}{\rho}}  \right )^{u-1} d_{\gamma_{u}}   \right).
\end{split}
\end{align}
By using the binomial expansion \cite{R171}, the expression  $\left (1-e^{-\frac{\gamma_u}{\rho}}  \right )^{u-1}$ in the above equation can be expanded as
\begin{align}
\label{eq:binomial_exp}
\left (1-e^{-\frac{\gamma_u}{\rho}}  \right )^{u-1} =\sum_{i=0}^{u-1}\binom{u-1}{i}(-1)^{i}e^{-\frac{\gamma_u}{\rho}i}.
\end{align}
Now, by taking some further mathematical simplification, we get
\begin{align}
\begin{split}
 \label{eq:app_u4}
&C_{e}^{u}=-\frac{1}{\theta_{u}{n}}{\rm{ln}}\left (\epsilon+\left ( 1-\epsilon \right )  \left (\frac{\xi_{u}}{\rho} \sum_{i=0}^{u-1}\binom{u-1}{i}(-1)^{i}  \underbrace{\int_{0}^{\infty}   \left ( 1+\alpha_u\gamma_{u} \right )^{2\zeta_{u}} e^{-\frac{\left ( V-u+1+i \right )}{\rho}\gamma_u} d_{\gamma_{u}} }_\text{$I_a$}  \right. \right. \\
&+\underbrace{\beta_{u} \int_{0}^{\infty}  \left (  1+\alpha_u\gamma_{u}  \right )^{2\zeta_{u}}   \sqrt{1-\left (  1+\alpha_u\gamma_{u}  \right )^{-2}}   e^{-\frac{\left ( V-u+1+i \right )}{\rho}\gamma_u}   d_{\gamma_{u}}}_\text{$I_b$} \\
&  \left. \left. +  \underbrace{ \frac{{\beta_{u}}^2}{2}  \int_{0}^{\infty} \left (  1+\alpha_u\gamma_{u}  \right )^{2\zeta_{u}}  \left (  1-\left (  1+\alpha_u\gamma_{u} \right )^{-2} \right )   e^{-\frac{\left ( V-u+1+i \right )}{\rho}\gamma_u}   d_{\gamma_{u}}}_\text{$I_c$}   \right)   \right).
\end{split}
\end{align}
Furthermore, we note that from (13.2.5) in \cite{R171},
\begin{align}
 \begin{split}
 \label{eq:app_u5}
{\rm{H}}\left ( a,b,z \right )=\frac{1}{\Gamma\left ( a \right )} \int_{0}^{\infty}e^{-zt}t^{a-1}\left ( 1+t \right )^{b-a-1} d_{t}, \ \ \rm{for} \ \rm{Re } \ a,\ \rm{Re} \ z>0,
\end{split}
\end{align}
where $\rm{H}(a,b,z)$ is the confluent hypergeometric function of the second kind and $\Gamma\left ( . \right )$  is the gamma function \cite{R171}.  Now, by applying  (\ref{eq:app_u5}) to  $I_a$, $I_b$ , and   $I_c$  of   (\ref{eq:app_u4}), we can have
\begin{align}
 \begin{split}
\label{eq:app_u6}
I_a={\rm{H}} \left ( 1,2+2\zeta_{u},{\frac{V-u+1+i }{\rho\alpha_{u}}} \right ),
\end{split}
\end{align}
\begin{align}
 \begin{split}
\label{eq:app_u7}
   I_b=\beta_u\left (    {\rm{H}} \left (1,2+2\zeta_{u},{\frac{ V-u+1+i }{\rho\alpha_{u}}} \right)  -\frac{1}{2}   {\rm{H}}\left (1,2\zeta_{u},{\frac{V-u+1+i }{\rho\alpha_{u}}}  \right)\right ),
\end{split}
\end{align}
\begin{align}
 \begin{split}
\label{eq:app_u8}
I_c=  \frac{{\beta_u}^2}{2}  \left (   {\rm{H}} \left(1,2+2\zeta_{u},{\frac{ V-u+1+i }{\rho\alpha_{u}}} \right )  -   {\rm{H}} \left (1,2\zeta_{u},{\frac{ V-u+1+i }{\rho\alpha_{u}}}  \right)\right ).
\end{split}
\end{align}
By inserting (\ref{eq:app_u6}), (\ref{eq:app_u7}), and (\ref{eq:app_u8}) into (\ref{eq:app_u4}) and using $K_u=\frac{\beta_{u}^{2}}{2}+\beta_u$ and $\eta_{u}=\frac{V-u+1+i}{\rho\alpha_u}$, we can  approximate the closed-form expression for $C_{e}^{u}$ as,
\begin{align}
 \begin{split}
\label{eq:app_u9}
C_{e}^{u}=-\frac{1}{\theta_{u}{n}}{\rm{ln}}\left ( \epsilon+\left ( 1-\epsilon \right )\left (\frac{\xi_{u}}{\rho\alpha_{u}} \sum_{i=0}^{u-1}\binom{u-1}{i}(-1)^{i} \right. \right. &  {\rm{H}} \left (1,2+2\zeta_{u},\eta_{u} \right) \left({\rm{K_u+1}} \right)  \\
&\left. \left.  -{\rm{H}} \left(1,2\zeta_{u},\eta_{u}  \right ) \left ({\rm{K_u}}-\frac{\beta_{u}}{2} \right)  \right )   \right ).
\end{split}
\end{align}
The closed-form expression for the achievable EC of strong user at high SNR presented as (\ref{eq:2ec_ub3}) can also be derived following the above steps.
\section*{APPENDIX B}
The achievable EC of weak user ($v_t$) with order statistics is formulated as
 \begin{align}
 \begin{split}
 \label{eq:app1ec_t5}
C_{e}^{t}=&-\frac{1}{\theta_{t}{n}}{\rm{ln}}\Bigg (\frac{\xi_{t}}{\rho} \int_{0}^{\infty} \Bigg (\epsilon+\left ( 1-\epsilon \right )
\left   (  \frac{\gamma_t+1 }{\alpha_u\gamma_t+1}   \right )^{2\zeta_{t}}+  \beta_{t}   \left (  \frac{\gamma_t+1 }{\alpha_u\gamma_t+1}   \right )^{2\zeta_{t}}   \\ &  \times   \sqrt{1-\left (  \frac{\gamma_t+1 }{\alpha_u\gamma_t+1}   \right )^{-2}}    +  \frac{ \beta_{t}   ^2}{2}  \left (  \frac{\gamma_t+1 }{\alpha_u\gamma_t+1}   \right  )^{2\zeta_{t}} \left (  1-\left ( \frac{\gamma_t+1 }{\alpha_u\gamma_t+1}   \right )^{-2}  \right )    \Bigg ) \\  &  \times     e^{-\frac{\left ( V-t+1 \right )\gamma_t}{\rho}}\left (1-e^{-\frac{\gamma_t}{\rho}}  \right )^{t-1}  d_{\gamma_{t}} \Bigg),
\end{split}
\end{align}
using the generalized binomial expansion \cite{R171}, we will expand the $\left ( \frac{\gamma_{t}+1 }{\alpha_u \gamma_{t}+1 }  \right )^{2\zeta_{t}}$ and $\left (1-e^{-\frac{\gamma_t}{\rho}}  \right )^{t-1} $ to simplify the (\ref{eq:app1ec_t5}), such as
\begin{align}
\begin{split}
\label{eq:app_t3}
\left ( \frac{\gamma_{t}+1 }{\alpha_u \gamma_{t}+1 }  \right )^{2\zeta_{t}}=\left (  \frac{1} {\alpha_u} \right )^{2\zeta_{t}}\left (  1+ \frac{\alpha_u-1}{\alpha_u \gamma_{t}+1 }  \right )^{2\zeta_{t}},
\end{split}
\end{align}
and then the $\left (  1+ \frac{\alpha_u+1}{\alpha_u \gamma_{t}+1 }  \right )^{2\zeta_{t}}$ can be expanded as,
\begin{align}
\begin{split}
\label{eq:app_t4}
\left (  1+ \frac{\alpha_u-1}{\alpha_u \gamma_{t}+1 }  \right )^{2\zeta_{t}}=\sum_{s=0}^{\infty}\binom{2\zeta_{t}}{s}\left (  \frac{\alpha_u-1}{\alpha_u \gamma_{t}+1}\right )^{s},
\end{split}
\end{align}
where from \cite{R171}, it is clear that
\begin{align}
 \begin{split}
 \label{eq:app_t4ii}
\left ( 1+a \right )^{x}=\sum^{\infty}_{y=0}\binom{x}{y}a^{y} \ \ \rm{for} \ \left | a \right |<1,
\end{split}
\end{align}
however for $y\geq 1$, $\binom{x}{y}$ can be,
\begin{align}
\begin{split}
\label{eq:app_t5}
\binom{x}{y}=\frac{x(x-1)...(x-y+1)}{y!}=\frac{(x)_y}{y!},
\end{split}
\end{align}
where $\binom{x}{0}=1$, and $(.)_y$ is the Pochhammer symbol. Using the binomial expansion, the expression  $\left (1-e^{-\frac{\gamma_t}{\rho}}  \right )^{t-1}$ from (\ref{eq:app1ec_t5}) can be expanded as
\begin{align}
\begin{split}
\label{eq:appbinomialt}
\left (1-e^{-\frac{\gamma_t}{\rho}}  \right )^{t-1}=\sum_{r=0}^{t-1}\binom{t-1}{r}(-1)^{r}e^{-\frac{\gamma_t}{\rho}r}.
\end{split}
\end{align}
By further simplifying the  (\ref{eq:app1ec_t5}) with the above expansions, we achieve
\begin{align}
\begin{split}
\label{eq:app_t6}
&C_{e}^{t}=-\frac{1}{\theta_{t}{n}}{\rm{ln}}\Bigg (\epsilon+\left ( 1-\epsilon \right ) \Bigg ( \frac{\alpha_{u}^{-2\zeta_{t}}\xi_{t} }
  {\rho} \Bigg( \sum_{r=0}^{t-1}\binom{t-1}{r} (-1)^{r}  \\ & \times \underbrace{\int_{0}^{\infty} \left (  1+2\zeta_{t}\frac{\alpha_{u}-1}{\alpha_{u}\gamma_{t}+1}+\sum_{s=2}^{\infty}\binom{2\zeta_{t}}{s}   \left (  \frac{\alpha_u-1}{\alpha_u \gamma_{t}+1}\right )^{s} \right )      e^{-\frac{(V-t+1+r)\gamma_t}{\rho}} d_{\gamma_t}}_\text{$I_1$} +\beta_t \sum_{r=0}^{t-1}\binom{t-1}{r} (-1)^{r}    \\   &  \times \underbrace{\int_{0}^{\infty}   \sqrt{1-\left (  \frac{\gamma_t+1 }{\alpha_u\gamma_t+1}   \right )^{-2}}  \left (  1+2\zeta_{t}\frac{\alpha_{u}-1}{\alpha_{u}\gamma_{t}+1}+\sum_{s=2}^{\infty}\binom{2\zeta_{t}}{s}\left (  \frac{\alpha_u-1}{\alpha_u \gamma_{t}+1}\right )^{s} \right ) e^{-\frac{(V-t+1+r)\gamma_t}{\rho}} d_{\gamma_t}}_\text{$I_2$} \\ & + \frac{{\beta_{t}}^2}{2} \sum_{r=0}^{t-1}\binom{t-1}{r} (-1)^{r}       \\  &    \times \underbrace{\int_{0}^{\infty}  \left (  1-\left ( \frac{\gamma_t+1 }{\alpha_u\gamma_t+1}   \right )^{-2}  \right )   \left (  1+2\zeta_{t}\frac{\alpha_{u}-1}{\alpha_{u}\gamma_{t}+1}+\sum_{s=2}^{\infty}\binom{2\zeta_{t}}{s}\left (  \frac{\alpha_u-1}{\alpha_u \gamma_{t}+1}\right )^{s} \right ) e^{-\frac{(V-t+1+r)\gamma_t}{\rho}} d_{\gamma_t}}_\text{$I_3$}    \Bigg )     \Bigg )\Bigg ).
\end{split}
\end{align}
For solving the above integrals $I_1$, $I_2$, and $I_3$, we use equations (3.353.2) and (3.352.4) from \cite{R172}, i.e.,
\begin{align}
\begin{split}
\label{eq:app_t7}
\int_{0}^{\infty}\frac{e^{-zt}}{t+b}d_{t}=-e^{bz}E_{i}(-bz), \ \left [\left | \rm{arg} \ b \right |<\pi, \ \rm{Re} \ z >0 \right ],
\end{split}
\end{align}
\begin{align}
\begin{split}
\label{eq:app_t8}
\int_{0}^{\infty}\frac{e^{-zt}}{(t+b)^n}d_{t}=\frac{1}{(n-1)!}\sum_{j=1}^{n-1} \left ( j-1 \right )!(-z)^{n-j-1} (b^{-j})
-\frac{(-z)^{n-1}}{(n-1)!}e^{bz}E_{i}(-bz), \\ \ \left [n\geq2, \ \left  | \rm{arg} \ b \right |<\pi, \ \rm{Re} \ z >0 \right ].
\end{split}
\end{align}
Finally, we can obtain the closed-form expression for $C_{e}^{t}$, as shown in (\ref{eq:1ec_t6}), by using (\ref{eq:app_t7}) and  (\ref{eq:app_t8}). The closed-form expression for the achievable EC of weak user at high SNR presented as (\ref{eq:2ec_tb3}) can also be derived following the above steps.


\begin{thebibliography}{10}
\bibliographystyle{IEEEtran}

\bibitem{R178}
W.~{Saad}, M.~{Bennis}, and M.~{Chen}, ``A vision of {6G} wireless systems:
  Applications, trends, technologies, and open research problems,'' \emph{IEEE
  Netw.}, to appear, 2019.

\bibitem{R136}
M.~{Bennis}, M.~{Debbah}, and H.~V. {Poor}, ``Ultra reliable and low-latency
  wireless communication: Tail, risk, and scale,'' \emph{Proc. IEEE}, vol. 106,
  no.~10, pp. 1834--1853, Oct. 2018.

\bibitem{R180}
Z.~{Ding}, X.~{Lei}, G.~K. {Karagiannidis}, R.~{Schober}, J.~{Yuan}, and V.~K.
  {Bhargava}, ``A survey on non-orthogonal multiple access for {5G} networks:
  Research challenges and future trends,'' \emph{IEEE J. Sel. Areas Commun.},
  vol.~35, no.~10, pp. 2181--2195, Oct. 2017.

\bibitem{R181}
S.~M.~R. {Islam}, N.~{Avazov}, O.~A. {Dobre}, and K.~{Kwak}, ``Power-domain
  non-orthogonal multiple access {(NOMA)} in {5G} systems: Potentials and
  challenges,'' \emph{IEEE Commun. Surveys Tuts.}, vol.~19, no.~2, pp.
  721--742, Oct. 2017.

\bibitem{R153}
G.~{Durisi}, T.~{Koch}, and P.~{Popovski}, ``Toward massive, ultra reliable,
  and low-latency wireless communication with short packets,'' \emph{Proc.
  IEEE}, vol. 104, no.~9, pp. 1711--1726, Sep. 2016.

\bibitem{R016}
P.~Popovski, J.~J. Nielsen, C.~Stefanovic, E.~De~Carvalho, E.~Strom, K.~F.
  Trillingsgaard, A.-S. Bana, D.~M. Kim, R.~Kotaba, J.~Park \emph{et~al.},
  ``Wireless access for ultra-reliable low-latency communication: Principles
  and building blocks,'' \emph{IEEE Netw.}, vol.~32, no.~2, pp. 16--23, Mar.
  2018.

\bibitem{R086}
Y.~{Gu}, H.~{Chen}, Y.~{Li}, and B.~{Vucetic}, ``Ultra-reliable short-packet
  communications: Half-duplex or full-duplex relaying?'' \emph{IEEE Wireless
  Commun. Lett.}, vol.~7, no.~3, pp. 348--351, Jun. 2018.

\bibitem{R139}
Y.~Polyanskiy, H.~V. Poor, and S.~Verd{\'u}, ``Channel coding rate in the
  finite blocklength regime,'' \emph{IEEE Trans. Inf. Theory}, vol.~56, no.~5,
  pp. 2307--2359, May 2010.

\bibitem{R047}
G.~Durisi, T.~Koch, J.~{\"O}stman, Y.~Polyanskiy, and W.~Yang, ``Short-packet
  communications over multiple-antenna {Rayleigh}-fading channels,'' \emph{IEEE
  Trans. Commun.}, vol.~64, no.~2, pp. 618--629, Feb. 2016.

\bibitem{R157}
J.~Choi, ``An effective capacity-based approach to multi-channel low-latency
  wireless communications,'' \emph{IEEE Trans. Commun.}, vol.~67, no.~3, pp.
  2476--2486, Nov. 2019.

\bibitem{R147}
D.~Wu and R.~Negi, ``Effective capacity: a wireless link model for support of
  quality of service,'' \emph{IEEE Trans. Wireless Commun.}, vol.~2, no.~4, pp.
  630--643, Jul. 2003.

\bibitem{R158}
M.~{Amjad}, L.~{Musavian}, and M.~H. {Rehmani}, ``Effective capacity in
  wireless networks: A comprehensive survey,'' \emph{IEEE Commun. Surveys
  Tuts.}, vol.~21, no.~4, pp. 3007--3038,, 2019.

\bibitem{R154}
M.~C. Gursoy, ``Throughput analysis of buffer-constrained wireless systems in
  the finite blocklength regime,'' \emph{EURASIP J. Wireless Commun. Net.},
  vol. 2013, no.~1, p. 290, Dec. 2013.

\bibitem{R156}
M.~Shehab, H.~Alves, and M.~Latva-aho, ``Effective capacity and power
  allocation for machine-type communication,'' \emph{IEEE Trans. Veh.
  Technol.}, vol.~68, no.~4, pp. 4098--4102, Apr. 2019.

\bibitem{R140}
Y.~Yu, H.~Chen, Y.~Li, Z.~Ding, and B.~Vucetic, ``On the performance of
  non-orthogonal multiple access in short-packet communications,'' \emph{IEEE
  Commun. Lett.}, vol.~22, no.~3, pp. 590--593, Mar. 2018.

\bibitem{R166}
X.~{Sun}, S.~{Yan}, N.~{Yang}, Z.~{Ding}, C.~{Shen}, and Z.~{Zhong},
  ``Short-packet downlink transmission with non-orthogonal multiple access,''
  \emph{IEEE Trans. Wireless Commun.}, vol.~17, no.~7, pp. 4550--4564, Jul.
  2018.

\bibitem{R176}
E.~Dosti, M.~Shehab, H.~Alves, and M.~Latva-aho, ``On the performance of
  non-orthogonal multiple access in the finite blocklength regime,'' \emph{Ad
  Hoc Netw.}, vol.~84, pp. 148--157, Mar. 2019.

\bibitem{R177}
Y.~Xu, C.~Shen, T.-H. Chang, S.-C. Lin, Y.~Zhao, and G.~Zhu, ``Energy-efficient
  non-orthogonal transmission under reliability and finite blocklength
  constraints,'' in \emph{IEEE Globecom Workshops (GC Wkshps)}, Marina Bay
  Sands, Singapore, Dec. 2017, pp. 1--6.

\bibitem{R159}
W.~Yu, L.~Musavian, and Q.~Ni, ``Link-layer capacity of {NOMA} under
  statistical delay {QoS} guarantees,'' \emph{IEEE Trans. Commun.}, vol.~66,
  no.~10, pp. 4907--4922, Oct. 2018.

\bibitem{R160}
M.~{Amjad} and L.~{Musavian}, ``Performance analysis of {NOMA} for
  ultra-reliable and low-latency communications,'' in \emph{IEEE Globecom
  Workshops (GC Wkshps)}, Abu Dhabi, Dec. 2018, pp. 1--5.

\bibitem{R183}
M.~{Amjad}, L.~{Musavian}, and S.~{Aissa}, ``{NOMA} versus {OMA} in finite
  blocklength regime: Link-layer rate performance,''
  \emph{arxiv.org/abs/1912.08119}, 2019.

\bibitem{R163}
H.~A. David and H.~N. Nagaraja, ``Order statistics,'' \emph{Encyclopedia of
  Statistical Sciences}, 2004.

\bibitem{R179}
J.~E. Gentle, \emph{Computational statistics}.\hskip 1em plus 0.5em minus
  0.4em\relax Springer, 2009, vol. 308.

\bibitem{R182}
C.~C. Maican, \emph{Integral Evaluations Using the Gamma and Beta Functions and
  Elliptic Integrals in Engineering: A Self-Study Approach}.\hskip 1em plus
  0.5em minus 0.4em\relax International Press, 2005.

\bibitem{R161}
C.-S. Chang, ``Stability, queue length, and delay of deterministic and
  stochastic queueing networks,'' \emph{IEEE Trans. Autom. Control}, vol.~39,
  no.~5, pp. 913--931, May 1994.

\bibitem{R167}
------, \emph{Performance guarantees in communication networks}.\hskip 1em plus
  0.5em minus 0.4em\relax Springer Science \& Business Media, 2012.

\bibitem{R168}
J.~Bucklew, \emph{Introduction to rare event simulation}.\hskip 1em plus 0.5em
  minus 0.4em\relax Springer Science \& Business Media, 2013.

\bibitem{R171}
M.~Abramowitz and I.~A. Stegun, ``Handbook of mathematical functions dover
  publications,'' \emph{New York}, p. 361, 1965.

\bibitem{R172}
I.~S. Gradshteyn and I.~M. Ryzhik, \emph{Table of integrals, series, and
  products}.\hskip 1em plus 0.5em minus 0.4em\relax Academic press, 2014.

\end{thebibliography}
\end{document}